\documentclass[10pt,journal,final]{IEEEtran}
\usepackage{amssymb,amsmath,graphicx,subfigure,enumerate,multirow}
\usepackage{citesort}
\usepackage{amsthm}
\usepackage{algorithm}
\usepackage{algorithmic}
\usepackage{stfloats}
 \allowdisplaybreaks
\newtheorem{theorem}{Theorem}
\newtheorem{cor}{Corollary}
\newtheorem{lemma}{Lemma}
\newtheorem{prop}{Proposition}
\newtheorem{remark}{Remark}
\newtheorem{assumption}{Assumption}

\IEEEoverridecommandlockouts

\begin{document}
\title{A Robust Statistics Approach\\to Minimum Variance Portfolio Optimization}
\author{
Liusha Yang\IEEEauthorrefmark{1}, Romain Couillet\IEEEauthorrefmark{2}, Matthew R. McKay\IEEEauthorrefmark{1}
\thanks{\IEEEauthorrefmark{1}L. Yang and M. R. McKay are with the Department of Electronic and Computer Engineering, Hong Kong University of Science and Technology, Clear Water Bay, Kowloon, Hong Kong. (email:lyangag@connect.ust.hk;eemckay@ust.hk).}
\thanks{\IEEEauthorrefmark{2}R. Couillet is with the Laboratoire de Signaux et Systmes (L2S, UMR8506), CNRS-CentraleSup\'{e}lec-Universit\'{e} Paris-Sud, 3 rue Joliot-Curie, 91192 Gif-sur-Yvette, France (email:romain.couillet@centralesupelec.fr).}
\thanks{Yang and McKay's work is supported by the Hong Kong Research Grants Council under grant number 16206914.
Couillet's work is supported by the ERC MORE EC--120133.}}
\maketitle

\begin{abstract}
We study the design of portfolios under a minimum risk criterion. The performance of the optimized portfolio relies on the accuracy of the estimated covariance matrix of the portfolio asset returns. For large portfolios, the number of available market returns is often of similar order to the number of assets, so that the sample covariance matrix performs poorly as a covariance estimator. Additionally, financial market data often contain outliers which, if not correctly handled, may further corrupt the covariance estimation. We address these shortcomings by studying the performance of a hybrid covariance matrix estimator based on Tyler's robust M-estimator and on Ledoit-Wolf's shrinkage estimator while assuming samples with heavy-tailed distribution. Employing recent results from random matrix theory, we develop a consistent estimator of (a scaled version of) the realized portfolio risk, which is minimized by optimizing online the shrinkage intensity.  Our portfolio optimization method is shown via simulations to outperform existing methods both for synthetic and real market data.
\end{abstract}

\section{Introduction}
The theory of portfolio optimization is generally associated
with the classical mean-variance optimization framework of Markowitz \cite{Markowitz1952}. The pitfalls of the mean-variance analysis are mainly related to its sensitivity to the estimation error of the means and covariance matrix of the asset returns.  It is nonetheless argued that estimates of the covariance matrix are more accurate than those of the expected returns \cite{merton1980estimating,Jagannathan2003}. Thus, many studies concentrate on improving the performance of the global minimum variance portfolio (GMVP), which provides the lowest possible portfolio risk and involves only the covariance matrix estimate.

The frequently used covariance estimator is the well-known sample covariance matrix (SCM). However, covariance estimates for portfolio optimization commonly involve few historical observations of sometimes up to a thousand assets. In such a case, the number of independent samples $n$ may be small compared to the covariance matrix dimension $N$, which suggests a poor performance of the SCM. The impact of the estimation error on the out-of-sample performance of the GMVP based on the SCM has already been analyzed in \cite{el2010high,karoui2013realized,bai2009enhancement,bouchaud2009financial}.

In the finance literature, several approaches have been proposed to get around the problem of the scarcity of samples. One approach is to impose some factor structure on the estimator of the covariance matrix \cite{chan1999portfolio,fan2008high}, which reduces the number of parameters to be estimated.
A second approach is to use as a covariance matrix estimator a weighted
average of the sample covariance matrix and another
estimator, such as the 1-factor covariance matrix or
the identity matrix \cite{ledoit2003improved,Ledoit&Wolf2004}. A third approach is a nonlinear shrinkage estimation approach \cite{ledoit2014nonlinear}, which modifies each eigenvalue of the SCM under the framework of Markowitz's portfolio selection.
A fourth approach comprises eigenvalue clipping methods \cite{laloux1999noise,laloux2000random,plerou2002random} whose underlying idea is to `clean' the SCM by filtering noisy eigenvalues claimed to convey little valuable information. This approach has also been employed recently in proposing novel vaccine design strategies for infectious diseases \cite{dahirel2011coordinate,quadeer2013statistical}, and its theoretical foundations have been examined in \cite{donoho2013optimal}. A fifth method employs a bootstrap-corrected estimator for the optimal return and its asset allocation, which reduces the error of over-prediction of the in-sample return by bootstrapping \cite{bai2009enhancement}. In contrast to all of these methods (which aim to improve the covariance matrix estimate), alternative methods have also been proposed which directly impose various constraints on the portfolio weights, such as a no-shortsale constraint \cite{Jagannathan2003}, a $L_1$ norm constraint and a $L_2$ norm constraint \cite{fan2012vast,demiguel2009generalized}. By bounding directly the portfolio-weight vector, it is demonstrated that the estimation error can be reduced, particularly when the portfolio size is large \cite{fan2012vast}.

In addition to the problem of sample deficiency, it is often the case that the return observations exhibit impulsiveness and local loss of stationarity \cite{cont2001empirical}, which is not addressed by the methods mentioned above and leads to performance degradation. The field of robust estimation \cite{huber1964robust,tyler1987distribution,maronna1976robust,kent1991redescending} intends to deal with this problem. However, classical robust covariance estimators generally require $n\gg N$ and do not perform well (or are not even defined) when $n\simeq N$, making them unsuitable for many modern applications. Recent works \cite{couillet2013robust,couillet2013random,zhang2014marchenko,pascal2013robustshr,abramovich2007diagonally,Hero2011,couillet2014large} based on random matrix theory have therefore considered robust estimation in the $n\simeq N$ regime. Two hybrid robust shrinkage covariance matrix estimates have been proposed in parallel in \cite{pascal2013robustshr,abramovich2007diagonally} and in \cite{Hero2011}, respectively, both of which estimators are built upon Tyler's robust M-estimatior  \cite{tyler1987distribution} and Ledoit-Wolf's shrinkage approach \cite{Ledoit&Wolf2004}. In \cite{couillet2014large}, the authors show, by means of random matrix theory, that in the large $n,N$ regime and under the assumption of elliptical vector observations, the estimators in \cite{pascal2013robustshr,abramovich2007diagonally} and \cite{Hero2011} perform essentially the same and can be analyzed thanks to their asymptotic closeness to well-known random matrix models. Therefore, in this paper, we concentrate on the estimator studied in \cite{pascal2013robustshr,abramovich2007diagonally}, which we denote by $\hat{\bf C}_{\rm ST}$ (ST standing for shrinkage Tyler). Namely,  for independent samples ${\bf x}_1,...,{\bf x}_n\in\mathbb{R}^N$ with zero mean, $\hat{\bf C}_{\rm ST}$ is the unique solution to the fixed-point equation $$\hat{\bf C}_{\rm ST}(\rho)=(1-\rho)\frac{1}{n}\sum_{t=1}^n\frac{{\bf x}_{t}{\bf x}_{t}^{T}}{\frac{1}{N}{\bf x}_{t}^T\hat{\bf C}_{\rm ST}^{-1}(\rho){\bf x}_{t}}+\rho{\bf I}_N$$ for any $\rho\in(\max\{1-\frac{N}{n},0\},1]$. It should be noted that the shrinkage structure even allows $n<N$.

This paper designs a novel minimum variance portfolio optimization strategy based on $\hat{\bf C}_{\rm ST}$ with a risk-minimizing (instead of Frobenius norm minimizing \cite{couillet2014large}) shrinkage parameter $\rho$. We first characterize the out-of-sample risk of the minimum variance portfolio with plug-in ST for all $\rho$ within a specified range. This is done by analyzing the uniform convergence of the achieved realized risk on $\rho$ in the double limit regime, where $N,n\rightarrow\infty$, with $c_N=N/n\rightarrow c\in(0,\infty)$. We subsequently provide a consistent estimator of the realized portfolio risk (or, more precisely, a scaled version of it) that is defined only in terms of the observed returns. Based on this, we obtain a risk-optimized ST covariance estimator by optimizing online over $\rho$, and thus our optimized portfolio.

The proposed portfolio selection is shown to achieve superior performance over the competing methods in \mbox{\cite{couillet2014large,Hero2011,Ledoit&Wolf2004,Fran2012}} in minimizing the realized portfolio risk under the GMVP framework for impulsive data.
The outperformance of our portfolio optimization strategy compared to other methods is demonstrated through Monte Carlo simulations with elliptically distributed samples, as well as with real data of historical (daily) stock returns from Hong Kong's Hang Seng Index (HSI).

\textit{Notations:} Boldface upper case letters denote matrices, boldface lower case letters denote column vectors, and standard lower case letters denote scalars. $(\cdot)^T$ denotes transpose. ${\bf I}_N$ denotes the $N\times N$ identity matrix and ${\bf 1}_N$ denotes an $N$-dimensional vector with all entries equal to one. ${\rm tr}[\cdot]$ denotes the matrix trace operator. $\mathbb{R}$ and $\mathbb{C}$ denote the real and complex fields of dimension specified by a superscript. $\|\cdot\|$ denotes the Euclidean norm for vectors and the spectral norm for matrices. The Dirac measure at point $x$ is denoted by ${\pmb\delta}_x$. The ordered eigenvalues of a symmetric matrix ${\bf X}$ of size $N\times N$ are denoted by $\lambda_1({\bf X})\leq...\leq\lambda_N(\bf X)$, and the cardinality of a set $\mathcal{C}\subset\mathbb{R}$ is denoted by $|\mathcal{C}|$. Letting ${\bf U},{\bf V}$ be symmetric $N\times N$ matrices, we write ${\bf U}\succcurlyeq{\bf V}$ if ${\bf U}-{\bf V}$ is positive semidefinite.

\section{Data model and problem formulation} \label{sec:proform}

 We consider a time series comprising ${\bf x}_1,...,{\bf x}_n\in\mathbb{R}^N$ logarithmic returns of $N$ financial assets. We assume the ${\bf x}_t$ to be independent and identically distributed (i.i.d.) with
\begin{align}\label{eq:timeseries}
{\bf x}_t={\pmb\mu}+\sqrt{\tau_t}{\bf C}_N^{1/2}{\bf y}_t, ~~~t=1,2,...,n,
\end{align}
where ${\pmb\mu}\in\mathbb{R}^N$ is the mean vector of the asset returns, ${\tau_t}$ is a real, positive random variable, ${\bf C}_N\in\mathbb{R}^{N\times N}$ is positive definite and ${\bf y}_t\in\mathbb{R}^N$ is a zero mean unitarily invariant random vector with norm $\|{\bf y}_t\|^2=N$, independent of the $\tau_i$'s. It is assumed that ${\pmb\mu}$ and ${\bf C}_N$ are time-invariant over the observation period.  Denote ${\bf z}_t={\bf C}_N^{1/2}{\bf y}_t$. The model (\ref{eq:timeseries}) for ${\bf x}_t$ embraces in particular the class of elliptical distributions, including the multivariate normal distribution, exponential distribution and the multivariate Student-T distribution as special cases. This model for ${\bf x}_t$ leads to tractable and adoptable design solutions and is a commonly used approximation of the impulsive nature of financial data \cite{ledoit2003improved}.

 Let ${\bf h}\in\mathbb{R}^N$ denote the portfolio selection, i.e., the vector of asset holdings in units of currency normalized by the total outstanding wealth, satisfying ${\bf h}^T{\bf 1}_N=1$. In this paper, short-selling is allowed, and thus the portfolio weights may be negative. Then the portfolio variance (or risk) over the investment period of interest is defined as $\sigma^2({\bf h})=E[|{\bf h}^T{\bf x}_t|^2]={\bf h}^T{\bf C}_N{\bf h}$ \cite{Markowitz1952}. Accordingly, the GMVP selection problem can be formulated as the following quadratic optimization problem with a linear constraint:
\begin{align}  \nonumber 
\min\limits_{\bf h}~~~\sigma^2({\bf h})~~~~~
{\rm s.t.}~~{\bf h}^T{\bf 1}_N=1.
\end{align}
This has the well-known solution $${\bf h}_{\rm GMVP}=\frac{{\bf C}_N^{-1}{\bf 1}_N}{{\bf 1}_N^T{\bf C}_N^{-1}{\bf 1}_N}$$
and the corresponding portfolio risk is
\begin{align} \label{eq:theoretical risk}
\sigma^2\left({\bf h}_{\rm GMVP}\right)=\frac{1}{{\bf 1}_N^T{\bf C}_N^{-1}{\bf 1}_N}.
\end{align}

Here, (\ref{eq:theoretical risk}) represents the theoretical minimum portfolio risk bound, attained upon knowing the covariance matrix ${\bf C}_N$ exactly. In practice, ${\bf C}_N$ is unknown, and instead we form an estimate, denoted by $\hat{\bf C}_N$. Thus, the GMVP selection based on the plug-in estimator $\hat{\bf C}_N$ is given by
\begin{align} \nonumber
\hat{\bf h}_{\rm GMVP}=\frac{\hat{\bf C}_N^{-1}{\bf 1}_N}{{\bf 1}_N^T\hat{\bf C}_N^{-1}{\bf 1}_N}.
\end{align}

The quality of $\hat{\bf h}_{\rm GMVP}$, implemented based on the in-sample covariance prediction $\hat{\bf C}_N$, can be measured by its achieved out-of-sample (or ``realized'') portfolio risk:
\begin{align} \nonumber
\sigma^2\left(\hat{\bf h}_{\rm GMVP}\right)=\frac{{\bf 1}_N^T\hat{\bf C}_N^{-1}{\bf C}_N\hat{\bf C}_N^{-1}{\bf 1}_N}{({\bf 1}_N^T\hat{\bf C}_N^{-1}{\bf 1}_N)^2}.
\end{align}
The goal is to construct a good estimator $\hat{\bf C}_N$, and consequently $\hat{\bf h}_{\rm GMVP}$, which minimizes this quantity.

Note that, for the naive uniform diversification rule, ${\bf h}=\frac{1}{N}{\bf 1}_N$. This is equivalent to setting $\hat{\bf C}_N={\bf I}_N$, and yields the realized portfolio risk: $\frac{{\bf 1}_N^T{\bf C}_N{\bf 1}_N}{N^2}$. Interestingly, this extremely simple strategy has been shown in \cite{DeMiguel2009} to outperform numerous optimized models and will serve as a benchmark in our work.

\section{Novel covariance estimator and portfolio design for minimizing risk} \label{sec:asympest}

\subsection{Tyler's robust M-estimator with linear shrinkage}
Consider the ST covariance matrix estimate introduced in \cite{pascal2013robustshr,abramovich2007diagonally}, built upon both Tyler's M-estimate \cite{tyler1987distribution} and the Ledoit-Wolf shrinkage estimator \cite{Ledoit&Wolf2004}.
This estimator accounts for the scarcity of samples, even allowing $N>n$, and exhibits robustness to outliers or impulsive samples, e.g., elliptically distributed data. It is defined as the unique solution to the following fixed-point equation for \mbox{$\rho\in(\max\{0,1-n/N\},1]$:}
\begin{align} \label{eq:SFPE}
\hat{\bf C}_{\rm ST}(\rho)=(1-\rho)\frac{1}{n}\sum_{t=1}^n\frac{\tilde{\bf x}_{t}\tilde{\bf x}_{t}^{T}}{\frac{1}{N}\tilde{\bf x}_{t}^T\hat{\bf C}_{\rm ST}^{-1}(\rho)\tilde{\bf x}_{t}}+\rho{\bf I}_N
\end{align}
where
$\tilde{\bf x}_{t}={\bf x}_{t}-\frac{1}{n}\sum_{i=1}^n{\bf x}_{i}$.

Since with probability one the ${\bf x}_t$ are linearly independent, $\hat{\bf C}_{\rm ST}(\rho)$ is almost surely defined for each $N$ and $n$ \cite[Theorem III.1]{pascal2013robustshr}.  The corresponding GMVP selection is
\begin{align} \nonumber
\hat{\bf h}_{\rm ST}(\rho)=\frac{\hat{\bf C}_{\rm ST}^{-1}(\rho){\bf 1}_N}{{\bf 1}_N^T\hat{\bf C}_{\rm ST}^{-1}(\rho){\bf 1}_N}
\end{align}
with realized portfolio risk
\begin{align} \label{eq:SFPE risk}
\sigma^2\left(\hat{\bf h}_{\rm ST}(\rho)\right)=\frac{{\bf 1}_N^T\hat{\bf C}_{\rm ST}^{-1}(\rho){\bf C}_N\hat{\bf C}_{\rm ST}^{-1}(\rho){\bf 1}_N}{({\bf 1}_N^T\hat{\bf C}_{\rm ST}^{-1}(\rho){\bf 1}_N)^2}.
\end{align}

Our goal is to optimize $\rho$ online such that (\ref{eq:SFPE risk}) is minimum. However, since (\ref{eq:SFPE risk}) involves ${\bf C}_N$ which is unobservable, this equation cannot be optimized directly. Also note that the naive approach of simply replacing ${\bf C}_N$ with $\hat{\bf C}_{\rm ST}(\rho)$ in (\ref{eq:SFPE risk}) would yield the so-called ``in-sample risk'', which underestimates the realized portfolio risk, leading to overly-optimistic investment decisions \cite{Fran2012}.
We tackle this problem by obtaining a consistent estimator for a scaled version of the realized risk (\ref{eq:SFPE risk}) as $n$ and $N$ go to infinity at the same rate. Contrary to classical asymptotic theory for time series analysis and mathematical statistics, which typically deal with the case of $N$ fixed and $n\rightarrow\infty$, a double-limiting condition is of more relevance for large portfolio problems, where $n$ is comparable to $N$. To this end, following \cite{Fran2012}, we first derive a deterministic asymptotic equivalent of (\ref{eq:SFPE risk}) and then provide a consistent estimator based on this.

\subsection{Deterministic equivalent of the realized portfolio risk}
For our asymptotic analysis, we assume the following:
\begin{assumption} \label{assump}
\hfill
\begin{enumerate}[a.]
\item As $N, n\rightarrow\infty$, $N/n=c_N\rightarrow c\in(0,\infty)$. \label{assump:1}
\item
    The ${\tau}_t,~t=1,...,n$ are i.i.d. $\tau_1,...,\tau_n\geq\xi$ a.s. for some $\xi>0$ and $E[\tau_1]<\infty$.\footnote{For technical reasons, made explicit in the appendix, we require the quantities $\tilde{\bf z}_{t}={\bf z}_t-\frac{1}{n}\sum_{i=1}^n{\bf z}_i\sqrt{\frac{\tau_i}{\tau_t}}$ to have controllable norms. This imposes the constraint $\tau_t\geq\xi>0$ which might be possible to relax at the expense of increased mathematical complexity.}\label{assump:2}
\item Denoting $0<\lambda_1\leq...\leq\lambda_N$ the ordered eigenvalues of ${\bf C}_N$,  as $N,n\rightarrow\infty$, $\nu_N\triangleq\frac{1}{N}\sum_{i=1}^N{\pmb\delta}_{\lambda_i}$ satisfies $\nu_N\rightarrow\nu$ weakly with $\nu\neq{\pmb\delta}_0$ almost everywhere. In addition, $\limsup_N\lambda_N<\infty$. \label{assump:3}
\end{enumerate}
\end{assumption}

We also introduce some further definitions, which will arise in our asymptotic analysis.
For $\rho\in(\max(0,1-c^{-1}),1]$, define
$\gamma$ the unique positive solution to
\begin{align} \label{def:gamma}
1=\int\dfrac{t}{\gamma\rho+(1-\rho)t}\nu(dt)
\end{align}
and
\begin{align} \nonumber
\beta=\int\frac{c\gamma^2t^2}{(\gamma\rho+(1-\rho)t)^2}\nu(dt).
\end{align}

The following theorem presents our first key result: a deterministic characterization of the asymptotic realized portfolio risk achieved with $\hat{\bf C}_{\rm ST}(\rho)$.
 \begin{theorem} \label{thrm:asymp}
 Let Assumption \ref{assump} hold. For $\varepsilon\in(0,\min\{1,c^{-1}\})$, define ${\mathcal{R}}_{\varepsilon}=[\varepsilon+\max\{0,1-c^{-1}\},1]$. Then, as $N,n\rightarrow\infty$,
 \begin{align} \label{eq:thrm1}
 \sup_{\rho\in{{\mathcal{R}}_{\varepsilon}}}\left|\sigma^2\left(\hat{\bf h}_{\rm ST}(\rho)\right)-\bar{\sigma}^2(\rho)\right|\stackrel{\rm a.s.}\longrightarrow0
 \end{align}
 where
\begin{align} \nonumber
 \bar{\sigma}^2(\rho)=&\frac{\gamma^2}{\gamma^2-\beta(1-\rho)^2} \times \\ \nonumber
 &\frac{{\bf 1}_N^T\!\left(\!\frac{1-\rho}{\gamma}{\bf C}_N+\rho{\bf I}_N\!\right)^{-1}\!{\bf C}_N\left(\!\frac{1-\rho}{\gamma}{\bf C}_N+\rho{\bf I}_N\!\right)^{-1}\!{\bf 1}_N}{\left({\bf 1}_N^T\left(\frac{1-\rho}{\gamma}{\bf C}_N+\rho{\bf I}_N\right)^{-1}{\bf 1}_N\right)^2}.
\end{align}
 \end{theorem}

 \emph{Proof:} See Appendix \ref{appx_prfth1}.


\begin{remark}
In Theorem \ref{thrm:asymp}, the set ${\mathcal{R}}_{\varepsilon}$ excludes the region $[0,\varepsilon+\max\{0,1-c^{-1}\})$. As we handle the uniformity of the convergence (\ref{eq:thrm1}), the proof of Theorem \ref{thrm:asymp} requires us to work on sequences $\left\{\rho_n\right\}_{n=1}^\infty$ of $\rho$. It is however difficult to handle the limit $\left|\sigma^2(\hat{\bf h}_{\rm ST}(\rho_n))-\bar{\sigma}^2(\rho_n)\right|$ for a sequence $\left\{\rho_n\right\}_{n=1}^\infty$ with $\rho_n\rightarrow0$. This follows from the same reasoning as that in \cite{couillet2014large} (see Equations (5) and (6) in Section 5.1 of \cite{couillet2014large} as well as Equation (\ref{eq:eplus}) in Appendix  \ref{appx:preliminaryresults} where $\rho_n\rightarrow\rho_0>0$ is necessary to ensure $e^+<1$). In the subsequent results, $\rho\in{\mathcal{R}}_{\varepsilon}$ is also required for the same reason.
\end{remark}

Theorem \ref{thrm:asymp} enables us to analyze the convergence of the realized portfolio risk in the regime of Assumption \ref{assump}-\ref{assump:1} for $\hat{\bf h}_{\rm ST}(\rho)$. In order to calibrate the shrinkage parameter $\rho$ for optimum GMVP performance, only the available sample data and certainly not the unknown ${\bf C}_N$ can be used. This is the objective of the subsequent section.

\subsection{Consistent estimation of scaled realized portfolio risk}
 Based on the observable data only, we can obtain an estimator of a scaled version of the realized portfolio risk, $\sigma^2(\hat{\bf h}_{\rm ST}(\rho))/\kappa$, where we define $\kappa\triangleq\int t\nu(dt)$.
We begin with the following lemma that provides a consistent estimator of $\gamma$, scaled by $1/\kappa$, which is denoted as $\hat{\gamma}_{\rm sc}$ (``sc'' standing for ``scaled'').
\begin{lemma} \label{lemma:gamma_est}
Under the settings of Theorem \ref{thrm:asymp}, as $N,n\rightarrow\infty$,
\begin{align} \label{eq:estgamma}
\sup_{\rho\in{{\mathcal{R}}_{\varepsilon}}}\left|{\hat \gamma}_{\rm sc}-{\gamma/\kappa}\right|\stackrel{\rm a.s.}\longrightarrow0
\end{align}
where
\begin{align} \nonumber
{\hat{\gamma}_{\rm sc}}=\dfrac{1}{1-(1-\rho)c_N}\dfrac{1}{n}\sum_{t=1}^n\frac{\tilde{\bf x}_{t}^T\hat{\bf C}_{\rm ST}^{-1}(\rho)\tilde{\bf x}_{t}}{\|\tilde{\bf x}_{t}\|^2}.
\end{align}
\end{lemma}

\emph{Proof:} See Appendix \ref{appx_prflemma1}.

%

The following theorem provides a consistent estimator of $\sigma^2(\hat{\bf h}_{\rm ST}(\rho))$, scaled by $1/\kappa$, which is denoted as $\hat{\sigma}_{\rm sc}^2(\rho)$. This is our second main result.

\begin{figure*}[!t]
\begin{align} \label{eq:estrisk}
 \hat{\sigma}_{\rm sc}^2(\rho)=\frac{\hat{\gamma}_{\rm sc}}{(1-\rho)-(1-\rho)^2c_N}
\frac{{\bf 1}_N^T\hat{\bf C}_{\rm ST}^{-1}(\rho)\left(\hat{\bf C}_{\rm ST}(\rho)-\rho{\bf I}_N\right)\hat{\bf C}_{\rm ST}^{-1}(\rho){\bf 1}_N}{({\bf 1}_N^T\hat{\bf C}_{\rm ST}^{-1}(\rho){\bf 1}_N)^2}.
\end{align}
\hrule
\end{figure*}
\begin{theorem} \label{thrm:estimate}
Under the settings of Theorem \ref{thrm:asymp}, as $N,n\rightarrow\infty$,
\begin{align}  \nonumber 
\sup_{\rho\in{{\mathcal{R}}_{\varepsilon}}}\left|\hat{\sigma}_{\rm sc}^2(\rho)-\frac{1}{\kappa}\sigma^2(\hat{\bf h}_{\rm ST}(\rho))\right|\stackrel{\rm a.s.}\longrightarrow0
\end{align}
where $\hat{\sigma}_{\rm sc}^2(\rho)$ is defined in (\ref{eq:estrisk}) at the top of the page.
\end{theorem}
\emph{Proof:} See Appendix \ref{appx_prfth2}.

Note that, since $\kappa$ is independent of $\rho$, the same $\rho$ minimizes both $\sigma^2(\hat{\bf h}_{\rm ST}(\rho))$ and $\sigma^2(\hat{\bf h}_{\rm ST}(\rho))/\kappa$.


The following corollary of Theorem \ref{thrm:estimate} is of fundamental importance, which demonstrates that choosing $\rho$ to minimize $\hat{\sigma}_{\rm sc}^2(\rho)$ is asymptotically equivalent to minimizing the unobservable $\sigma^2(\hat{\bf h}_{\rm ST}(\rho))$.
\begin{cor} \label{cor:minrho}
Denote ${\rho}^o$ and $\rho^*$ the minimizers of $\hat{\sigma}_{\rm sc}^2(\rho)$ and $\sigma^2(\hat{\bf h}_{\rm ST}(\rho))$ over ${\mathcal{R}}_{\varepsilon}$, respectively. Then, under the settings of Theorem \ref{thrm:asymp} and Theorem \ref{thrm:estimate}, as $N,n\rightarrow\infty$,
\begin{align} \nonumber 
\left|\sigma^2\left(\hat{\bf h}_{\rm ST}\left({\rho}^o\right)\right)-\sigma^2\left(\hat{\bf h}_{\rm ST}\left(\rho^*\right)\right)\right|\stackrel{\rm a.s.}\longrightarrow0.
\end{align}
\end{cor}
\emph{Proof:}  See Appendix \ref{appx_prfcor1}.

With this result, the GMVP optimization problem is now reduced to the minimization of $\hat{\sigma}_{\rm sc}^2(\rho)$, which can be done with a simple numerical search.

To summarise, given $n$ past return observations of $N$ assets, our proposed algorithm to construct a portfolio with minimal risk can be described as follows:
\begin{algorithm}[htb]
\caption{Proposed algorithm for GMVP optimization}
\begin{enumerate}
\item Compute the optimized shrinkage parameter via a numerical search
\begin{align}
\nonumber&{\rho}^o=\mathop{\arg\min}_{\rho\in[\varepsilon+\max\{0,1-c_N^{-1}\},1]}\left\{\hat{\sigma}_{\rm sc}^2(\rho)\right\}.
\end{align}
\item Form the risk-minimizing ST estimator $\hat{\bf C}_{\rm ST}^o$, the unique solution to
\begin{align}\nonumber
\hat{\bf C}_{\rm ST}^o= (1-{\rho}^o)\dfrac{1}{n}\sum_{t=1}^n\dfrac{\tilde{\bf x}_{t}\tilde{\bf x}_{t}^T}{\frac{1}{N}\tilde{\bf x}_{t}^T{\hat{\bf C}_{\rm ST}}^{o-1}\tilde{\bf x}_{t}}+{\rho}^o{\bf I}_N.
\end{align}
\item Construct the optimized portfolio
\begin{align} \nonumber
\hat{\bf h}_{{\rm ST}}^o=\frac{\hat{\bf C}_{\rm ST}^{o-1}{\bf 1}_N}{{\bf 1}_N^T\hat{\bf C}_{\rm ST}^{o-1}{\bf 1}_N}.
\end{align}
\end{enumerate}
\end{algorithm}

\section{Simulation results} \label{sec:sim}

We use both synthetic data and real market data to show the performance of $\hat{\bf C}_{\rm ST}^o$ compared to the following competing methods:
\begin{enumerate}
\item $\hat{\bf C}_{\rm P}$, referred to as the Abramovich-Pascal estimate from \cite{couillet2014large};
\item  $ \hat{\bf C}_{\rm C}$, referred to as the Chen estimate from \cite{couillet2014large};
\item $\hat{\bf C}_{\rm C2}$, the oracle estimator in \cite{Hero2011}, which has the same structure as $\hat{\bf C}_{\rm C}$, but resorts to solving an approximate problem of minimizing the Frobenius distance to find the optimal shrinkage;
\item $\hat{\bf C}_{\rm LW}$, the Ledoit-Wolf shrinkage estimator in \cite{Ledoit&Wolf2004};
\item $\hat{\bf C}_{\rm R}$, the Rubio estimator proposed in \cite{Fran2012}, which has the same structure as $\hat{\bf C}_{\rm LW}$, but with $\rho$ calibrated based on the GMVP framework, as in the present article.
\end{enumerate}

\subsection{Synthetic data simulations}
The synthetic data are generated i.i.d.\@ from a multivariate Student-T distribution, where $\sqrt{\tau_t}=\sqrt{d/\chi_d^2}$, $d=3$ and $\chi_d^2$ is a Chi-square random variable with $d$ degree of freedom. We set $N=200$. The mean vector ${\pmb\mu}$ can be set arbitrarily since it is discarded by the empirical mean, having no impact on the covariance estimates.
We assume the population covariance matrix ${\bf C}_N$ is based on a one-factor return structure \cite{mackinlay2000asset}:  ${\bf C}_N={\bf b}{\bf b}^T\sigma^2+{\bf\Sigma}$,
where $\sigma=0.16$. The factor loadings ${\bf b}\in\mathbb{R}^N$ are evenly spread between $0.5$ and $1.5$. The residual variance matrix ${\bf\Sigma}\in\mathbb{R}^{N\times N}$ is set to be diagonal and proportional to the identity matrix: ${\bf\Sigma}=\sigma_r^2{\bf I}$, where $\sigma_r=0.2$.

\begin{figure}[htb]
\centering
\includegraphics[width=0.9\linewidth]{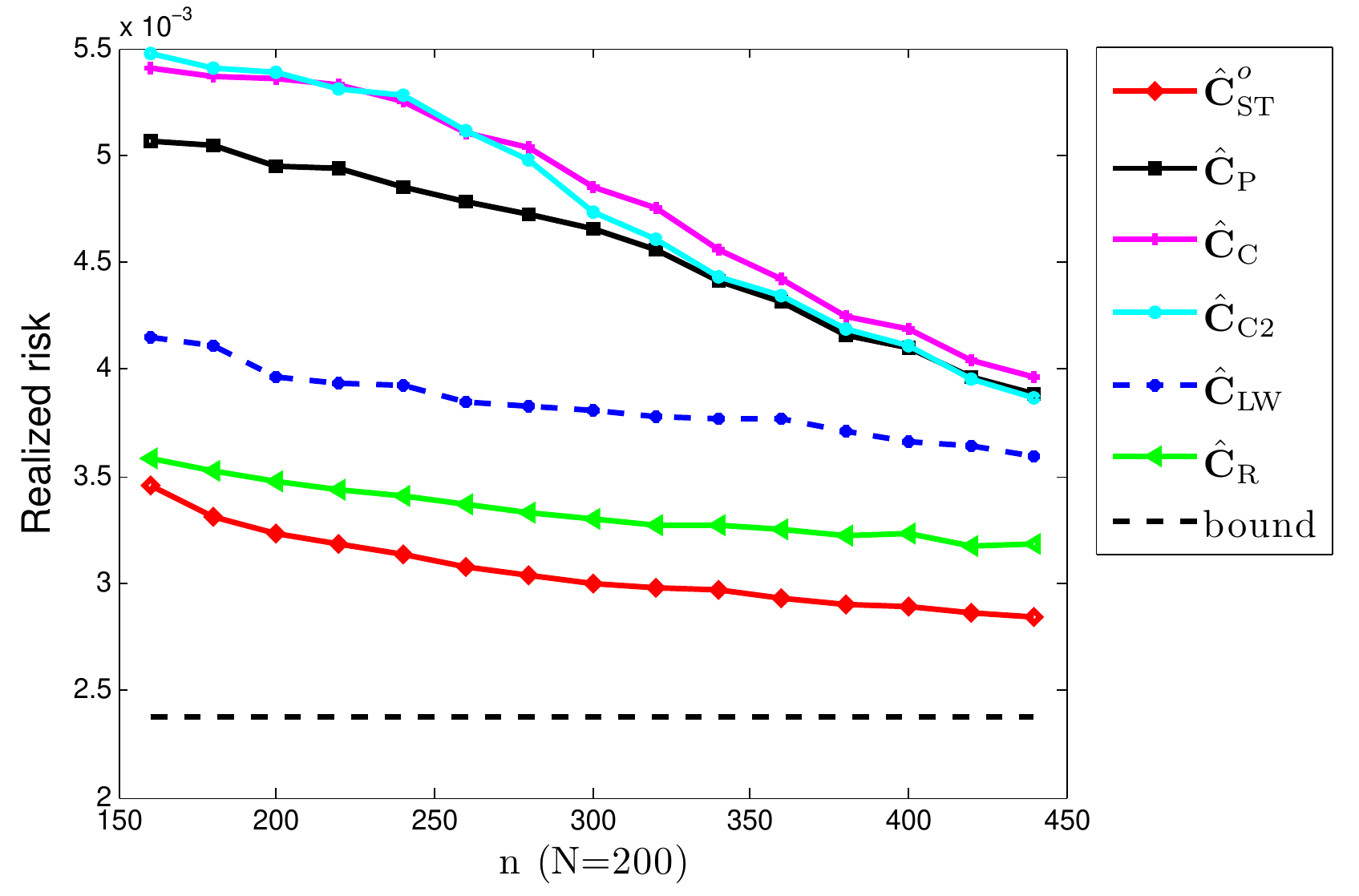}
\caption{The average realized portfolio risk of different covariance estimators in the GMVP framework using synthetic data. }
\label{fig:risk_N200_onefactor}
\end{figure}

Fig.\;\ref{fig:risk_N200_onefactor} illustrates the performance of different estimation approaches in terms of the realized risk, averaged over $200$ Monte Carlo simulations. The risk bound is computed by (\ref{eq:theoretical risk}), the theoretical minimum portfolio risk. Compared to other methods, our proposed estimator $\hat{\bf C}_{\rm ST}^o$ achieves the smallest realized risk for both $n\leq N$ and $n>N$. We omit the realized risks achieved by $\hat{\bf C}_N={\bf I}_N$ as they are uniformly more than five times as large as those achieved by the other methods.

\begin{figure}[htb]
\centering
\includegraphics[width=0.9\linewidth]{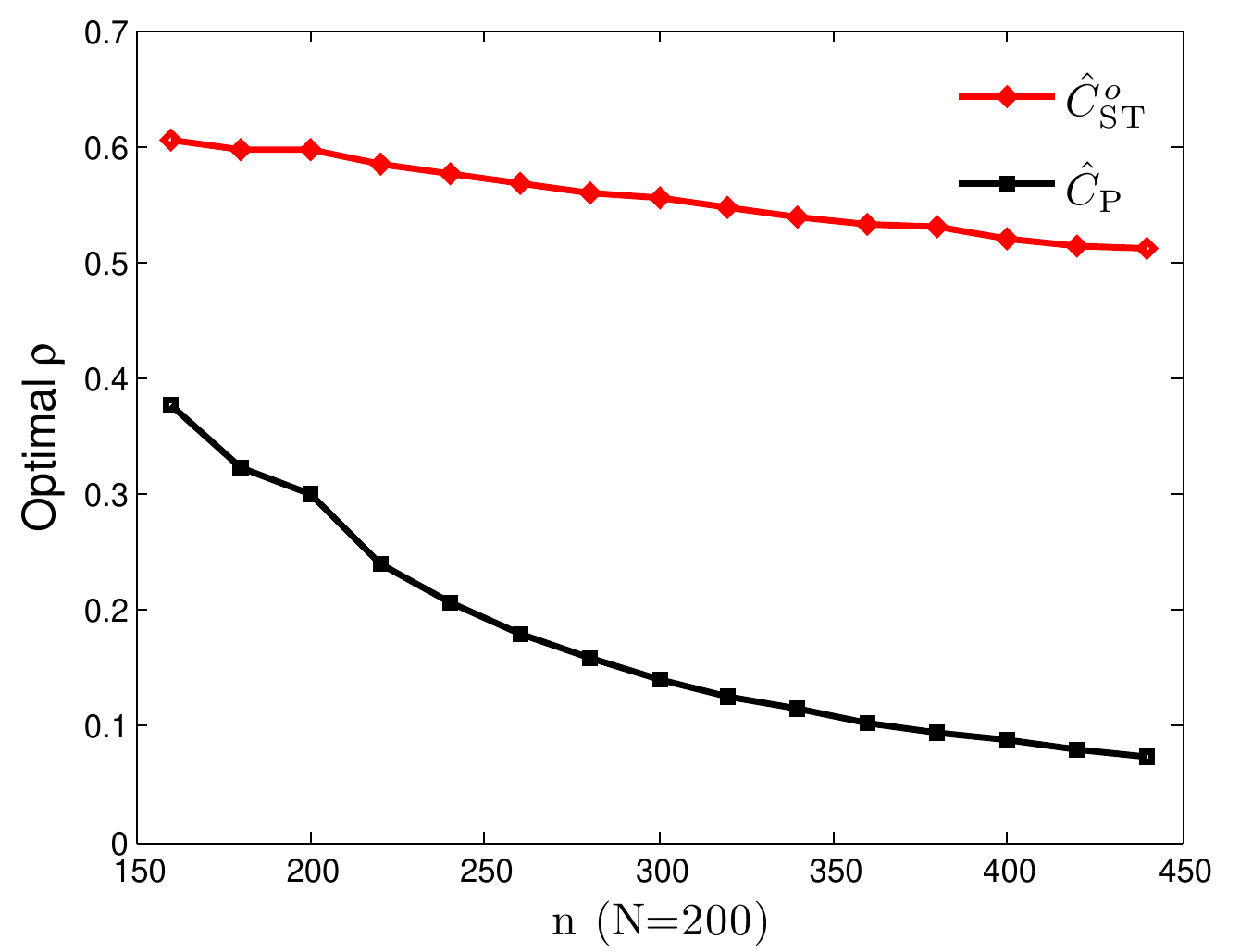}
\caption{The optimal shrinkage parameters of $\hat{\bf C}_{\rm ST}^o$ and $\hat{\bf C}_{\rm P}$ in the synthetic data simulation. }
\label{fig:rho_risk_N200_onefactor}
\end{figure}

It is interesting to compare the optimized $\rho$ of $\hat{\bf C}_{\rm ST}^o$ and $\hat{\bf C}_{\rm P}$. They are both solutions of (\ref{eq:SFPE}), but with $\rho$ optimized under different metrics: minimizing the risk and minimizing the Frobenius distance, respectively. As shown in Fig.\;\ref{fig:rho_risk_N200_onefactor}, the optimal shrinkage parameter varies under different metrics. Interestingly, optimizing $\rho$ under the risk function as opposed to the Frobenius distance leads to more aggressive shrinkage (regularization) towards the identity matrix, thus producing a portfolio allocation which is closer to the uniform allocation policy.

\subsection{Real market data simulations}
We now investigate the out-of-sample portfolio performance of the different estimators with the real market data. We consider the stocks comprising the HSI. In particular, we use the dividend-adjusted daily closing prices downloaded from the Yahoo Finance database to obtain the continuously compounded (logarithmic) returns for the $45$ constituents of the HSI over $L=736$ working days, from Jan. $3$, $2011$ to Dec. $31$, $2013$ (excluding weekends and public holidays).

As conventionally done in the financial literature, the out-of-sample evaluation is defined in terms of a rolling window method. At a particular day $t$, we use the previous $n$ days (i.e., from $t-n$ to $t-1$) as the training window for covariance estimation and construct the portfolio selection $\hat{\bf h}_{\rm GMVP}$.  We then use $\hat{\bf h}_{\rm GMVP}$ to compute the portfolio returns in the following $10$ days. Next the window is shifted $10$ days forward and the portfolio returns for another $10$ days are computed. This procedure is repeated until the end of the data. The realized risk is computed conventionally as the annualized sample standard deviation of the corresponding GMVP returns. In our tests, different training window lengths are considered.

\begin{figure}[htb]
\centering
\includegraphics[width=0.9\linewidth]{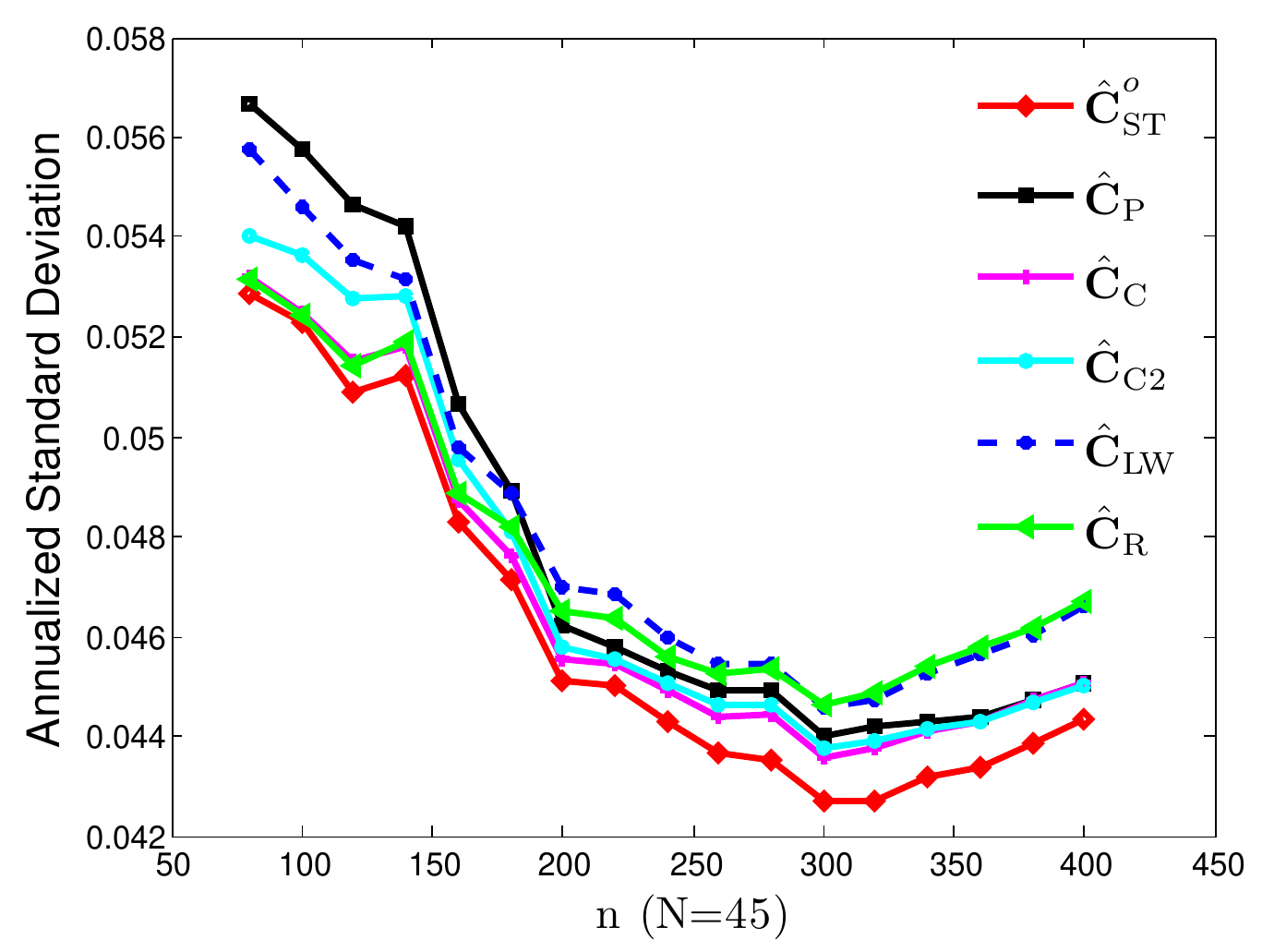}
\caption{Realized portfolio risks achieved out-of-sample over $736$ days of HSI real market data (from Jan. $3$, $2011$ to Dec. $31$, $2013$)
by a GMVP implemented using different covariance estimators. }
\label{fig:risk_HSI_N45}
\end{figure}

Fig. \ref{fig:risk_HSI_N45} shows that the proposed $\hat{\bf C}_{\rm ST}^o$ achieves the smallest realized risk. It outperforms the other methods over the entire span of considered estimation windows. The realized risk achieved by $\hat{\bf C}_N={\bf I}_N$ is also omitted here because it is more than double as those achieved by the competing methods. When the estimation window is too long (e.g., greater than $320$ days), we observe that the performance starts to systematically degrade. This is presumably due to a lack of stationarity in the data over such long durations. This highlights an interesting phenomenon worthy of further consideration, but a detailed study falls beyond the scope of the current contribution.

When the estimation window length is $300$, the lowest risk is achieved by $\hat{\bf C}_{\rm ST}^o$.  Table \ref{tab: data_45} presents the risks obtained by the different covariance estimators at the optimal estimation window length of $300$. We also test whether the pairwise differences between the portfolio variance achieved by $\hat{\bf C}_{\rm ST}^o$ and each benchmark strategy are statistically different from zero. Since standard hypothesis tests are not valid when returns have tails heavier than the normal distribution or are correlated across time,
we follow the method described in \cite{ledoit2008robust} and \cite{Ledoit2011robust} and employ a studentized version of the circular block bootstrap \cite{politis1994stationary} to do the test. The $p$-values are computed under the null hypothesis that the portfolio variance achieved by a particular benchmark covariance matrix estimator is equal to that achieved by $\hat{\bf C}_{\rm ST}^o$. We use a block length $b=5$ and base our reported $p$-values on $2000$ bootstrap iterations. We also compute the $p$-values when the block lengths are $b=1$ and $b=10$. The interpretation of the results does not change for $b=1$, $b=5$, or $b=10$. This implies that the temporal correlations of the stock returns are weak and our i.i.d. assumption on the data is acceptable. In the row reporting the risks, statistically significant outperformance of $\hat{\bf C}_{\rm ST}^o$ over other methods is denoted by asterisks: ** denotes significance at the $0.01$ level ($p<0.01$) and * denotes significance at the $0.05$ level ($p<0.05$). It can be seen from Table \ref{tab: data_45} that the outperformance of our proposed method is statistically significant, with $p<0.05$ in all cases.

\begin{figure}[htb]
\centering
\includegraphics[width=0.9\linewidth]{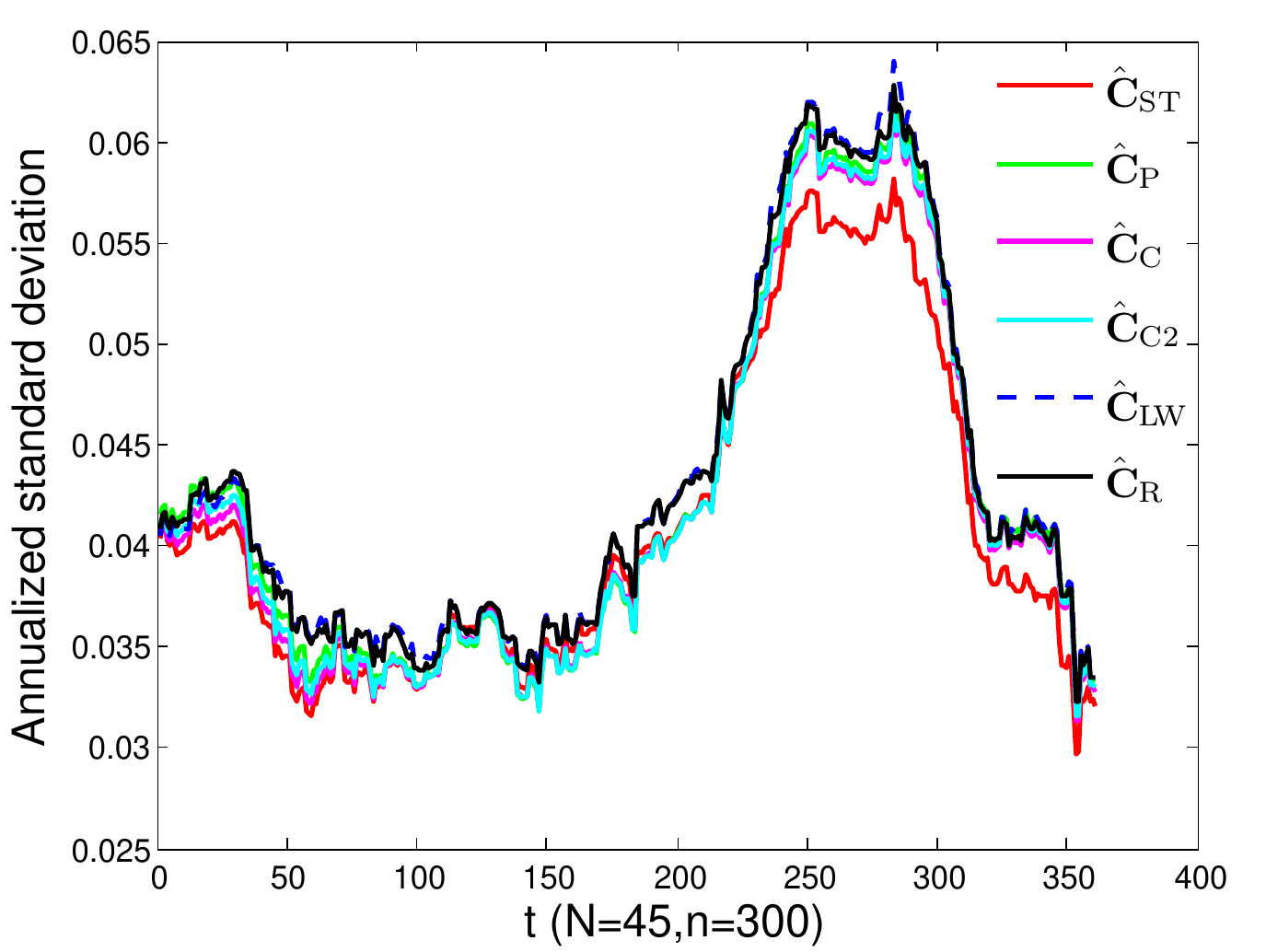}
\caption{Annualized rolling-window standard deviations of the most recent $70$ out-of-sample log returns for the GMVP based on different covariance matrix estimators.}
\label{fig:risk_changewithtime_HSI_n300_70}
\end{figure}

As a further comparison to investigate the performance with finer temporal resolution than that in Fig. \ref{fig:risk_HSI_N45}, we carry out a rolling-window analysis on the realized risks. Under the optimal estimation window length of $300$, we obtain $436$ out-of-sample portfolio returns. From the start of the data, we use the most recent $70$ out-of-sample portfolio returns to compute the (annualized) standard deviations of the GMVP. Shifting one day forward, we repeat this procedure until the end of the portfolio returns. For each covariance matrix estimator, this results in $367$ risk measurements, which are then displayed in a time series plot, Fig. \ref{fig:risk_changewithtime_HSI_n300_70}.  We find that $69.2\%$ of the time, $\hat{\bf C}_{\rm ST}^o$ achieves the lowest risk among all alternative methods. In addition, during the period of high volatility, that is, when $230<t<300$, $\hat{\bf C}_{\rm ST}^o$ exhibits the greatest outperformance. This justifies that our proposed GMVP optimization strategy is robust to market fluctuations and even possibly to outliers.


\begin{table*}[!t]\caption{Realized portfolio risks (annualized standard deviations) and the corresponding p-values under different covariance matrix estimators.}
\label{tab: data_45}
\begin{center}
\begin{tabular}{|c|c|c|c|c|c|c|c|c|}%
\hline
  Dataset&Statistic& $\hat{\bf C}_{\rm ST}^o$ & $\hat{\bf C}_{\rm P}$ & $\hat{\bf C}_{\rm C}$ & $\hat{\bf C}_{\rm C2}$ & $\hat{\bf C}_{\rm LW}$	& $\hat{\bf C}_{\rm R}$& ${\bf I}_N$ \\\hline
\multirow{2}{*}{HSI}&Risk (n=300) & $0.0419$	& $0.0433^{**}$	& $0.0428^{*}$	& $0.0430^{*}$ &	$0.0438^{**}$	& $0.0439^{**}$ & $0.1112^{**}$\\ \cline{2-9}
&$p$-value  & $1.000$	& $0.009$	& $0.028$	& $0.041$	& $0.001$	& $0.001$ & $0.000$ \\\hline
\end{tabular}
\end{center}
\end{table*}

\section{Conclusions} \label{sec:con}
We have proposed a novel minimum-variance portfolio optimization strategy based on a robust shrinkage covariance estimator with a shrinkage parameter calibrated to minimize the realized portfolio risk. Our strategy has been shown to be robust to finite-sampling effects as well as to the impulsive characteristics of the data. It has been demonstrated that our approach outperforms more standard techniques in terms of the realized portfolio risk, both for synthetic data and for real historical stock returns from Hong Kong's HSI.
Although we base our analysis on the assumption of the absence of the outliers, a recent study \cite{morales2015large} has shown that the robust covariance estimator $\hat{\bf C}_{\rm ST}$ is resilient to arbitrary outliers by appropriately weighting good versus outlying data. This is somewhat confirmed by our real data tests and is worth investigating further.


Even though GMVP is not an optimal portfolio in terms of the Sharpe ratio or return maximization at a given level of risk, many empirical studies \cite{jorion1991bayesian,chopra1993effect} has shown that an investment in the GMVP often yields better out-of-sample results than other mean-variance portfolios, because of the poor estimates of the means of the asset returns. Therefore, besides the robust estimation of the covariance matrix, it would be of interest to take into account the robust estimation of the means and further develop robust approaches to the various portfolio optimization strategies that involve both the estimates of the means and the covariance matrix of the asset returns, such as Sharpe ratio maximization or Markowitz's mean-variance portfolio optimization. These considerations are left to future work.
 \begin{appendices}
\section{Preliminary results} \label{appx:preliminaryresults}
In this appendix we provide some preparatory lemmas that are essential for the proof of the main theorems. 
From now on, for readability, we discard all unnecessary indices $\rho$ when no confusion is possible.

We start by rewriting $\hat{\bf C}_{\rm ST}$ in a more convenient form.
%
%
Denoting
\begin{align} \nonumber
\tilde{\bf z}_{t}={\bf z}_t-\frac{1}{n}{\bf Z}_N\sqrt{\frac{\pmb \tau}{\tau_t}},~~~~t=1,2,...,n,
\end{align}
with $\sqrt{\pmb\tau}=(\sqrt{\tau_1},...,\sqrt{\tau_n})^T$ and ${\bf Z}_N=[{\bf z}_1,...,{\bf z}_n]$, after some basic algebra, we obtain
\begin{align} \nonumber
\hat{\bf C}_{\rm ST}=(1-\rho)\frac{1}{n}\sum_{t=1}^n\frac{\tilde{\bf z}_{t}\tilde{\bf z}_{t}^{T}}{\frac{1}{N}\tilde{\bf z}_{t}^T\hat{\bf C}_{\rm ST}^{-1}\tilde{\bf z}_{t}}+\rho{\bf I}_N.
\end{align}

Denoting $\hat{\bf C}_{(t)}\triangleq\hat{\bf C}_{\rm ST}-(1-\rho)\frac{1}{n}\frac{\tilde{\bf z}_{t}\tilde{\bf z}_{t}^{T}}{\frac{1}{N}\tilde{\bf z}_{t}^T\hat{\bf C}_{\rm ST}^{-1}\tilde{\bf z}_{t}}$ and using $({\bf A}+r{\pmb\upsilon}{\pmb\upsilon}^T)^{-1}{\pmb\upsilon}={\bf A}^{-1}{\pmb\upsilon}/(1+r{\pmb\upsilon}^T{\bf A}^{-1}{\pmb\upsilon})$ for positive definite matrix ${\bf A}$, vector ${\pmb\upsilon}$ and scalar $r>0$, we have
\begin{align} \nonumber
\frac{1}{N}\tilde{\bf z}_{t}^T\hat{\bf C}_{\rm ST}^{-1}\tilde{\bf z}_{t}=\frac{\frac{1}{N}\tilde{\bf z}_{t}^T\hat{\bf C}_{(t)}^{-1}\tilde{\bf z}_{t}}{1+(1-\rho)c_N\frac{\frac{1}{N}\tilde{\bf z}_{t}^T\hat{\bf C}_{(t)}^{-1}\tilde{\bf z}_{t}}{\frac{1}{N}\tilde{\bf z}_{t}^T\hat{\bf C}_{\rm ST}^{-1}\tilde{\bf z}_{t}}}
\end{align}
so that
\begin{align} \label{eq:change}
\frac{1}{N}\tilde{\bf z}_{t}^T\hat{\bf C}_{\rm ST}^{-1}\tilde{\bf z}_{t}=(1-(1-\rho)c_N)\frac{1}{N}\tilde{\bf z}_{t}^T\hat{\bf C}_{(t)}^{-1}\tilde{\bf z}_{t}
\end{align}
and we can rewrite $\hat{\bf C}_{\rm ST}$ as
\begin{align} \nonumber
\hat{\bf C}_{\rm ST}=\frac{1-\rho}{1-(1-\rho)c_N}\frac{1}{n}\sum_{t=1}^n\frac{\tilde{\bf z}_{t}\tilde{\bf z}_{t}^{T}}{\frac{1}{N}\tilde{\bf z}_{t}^T\hat{\bf C}_{(t)}^{-1}\tilde{\bf z}_{t}}+\rho{\bf I}_N.
\end{align}

For $t\in\{1,...,n\}$, denote $\hat{d}_t(\rho)\triangleq\frac{1}{N}\tilde{\bf z}_{t}^T\hat{\bf C}_{(t)}^{-1}\tilde{\bf z}_{t}$. The following lemma gives a deterministic approximation of $\hat{d}_t(\rho)$, which later helps to show that, up to scaling, $\hat{\bf C}_{\rm ST}$ is somewhat similar to $\sum_{t=1}^n{\bf z}_{t}{\bf z}_{t}^T$, which is not observable.
\begin{lemma} \label{lemma:dgamma}
Under the settings of Theorem \ref{thrm:asymp}, as $N,n\rightarrow\infty$,
\begin{align} \nonumber
\sup_{\rho\in{\mathcal{R}}_\varepsilon}\max_{1\leq t\leq n}\left|\hat{d}_t(\rho)-{\gamma}(\rho)\right|\stackrel{\rm a.s.}\longrightarrow0.
\end{align}
\end{lemma}
\emph{Proof:}
This is proved via a contradiction argument, which follows along lines similar to the proof in \cite{couillet2014large}. The main difference lies in that we re-center the sample data by subtracting the sample mean, while the samples are assumed to be zero mean in \cite{couillet2014large}. By subtracting the sample mean, the re-centered data are correlated and some $\sqrt{\tau_t}$ terms still remain in $\hat{\bf C}_{\rm ST}$, which introduces new technical difficulties.

Assuming (by relabelling) that $\hat{d}_1(\rho)\leq...\leq\hat{d}_n(\rho)$, we first prove that for any fixed $\ell>0$, $\hat{d}_{{n}}(\rho)$ is bounded above by ${\gamma}(\rho)+\ell$ for all large $n$, uniformly on $\rho\in\mathcal{R}_\varepsilon$.
Since ${\bf U}\succcurlyeq{\bf V}\Rightarrow{\bf V}^{-1}\succcurlyeq{\bf U}^{-1}$, for positive definite matrices ${\bf U}$ and ${\bf V}$, we obtain
\begin{align} \nonumber 
\hat{d}_{{n}}(\rho)&=\frac{1}{N}\tilde{\bf z}_{{n}}^T\left(\frac{1-\rho}{1-(1-\rho)c_N}\frac{1}{n} \sum_{t=1}^{n-1}\frac{\tilde{\bf z}_{t}\tilde{\bf z}_{t}^T}{\hat{d}_t(\rho)}+\rho{\bf I}_N\right)^{-1}\tilde{\bf z}_{{n}} \\ \nonumber
&\leq\frac{1}{N}\tilde{\bf z}_{{n}}^T\left(\frac{1-\rho}{1-(1-\rho)c_N}\frac{1}{n} \sum_{t=1}^{n-1}\frac{\tilde{\bf z}_{t}\tilde{\bf z}_{t}^T}{\hat{d}_{{n}}(\rho)}+\rho{\bf I}_N\right)^{-1}\tilde{\bf z}_{{n}}.
\end{align}
Since $\tilde{\bf z}_{{n}}\neq 0$ with probability one, this implies
\begin{align} \label{eq:neq_d_n}
1\leq\frac{1}{N}\tilde{\bf z}_{{n}}^T\left(\frac{1-\rho}{1-(1-\rho)c_N}\frac{1}{n} \sum_{t=1}^{n-1}\tilde{\bf z}_{t}\tilde{\bf z}_{t}^T+\hat{d}_{{n}}(\rho)\rho{\bf I}_N\right)^{-1}\tilde{\bf z}_{{n}}.
\end{align}

Assume that there exists a sequence $\{\rho_n\}_{n=1}^\infty$ over which $\hat{d}_{{n}}(\rho_n)>{\gamma}(\rho_n)+\ell$ infinitely often, for some fixed $\ell>0$. Since $\{\rho_n\}_{n=1}^\infty$ is bounded, it has a limit point $\rho_0\in{\mathcal{R}}_\varepsilon$. Let us restrict ourselves to such a subsequence on which $\rho_n\rightarrow\rho_0>0$ and $\hat{d}_{{n}}(\rho_n)>{\gamma}(\rho_n)+\ell$. On this subsequence, from (\ref{eq:neq_d_n}), we have $\widetilde{m}_{N,n}\geq1$, where $\widetilde{m}_{N,n}=\frac{1}{N}\tilde{\bf z}_j^T\widetilde{\bf M}_{N,n}\tilde{\bf z}_n$ and $\widetilde{\bf M}_{N,n}=\left(\frac{1-\rho_n}{1-(1-\rho_n)c_N}\frac{1}{n} \sum_{t=1 }^n\tilde{\bf z}_{t}\tilde{\bf z}_{t}^T\!+({\gamma}(\rho_n)+\ell)\rho_n{\bf I}_N\right)^{-1}$.

The quadratic form $\widetilde{m}_{N,j}$ is amenable to large random matrix analysis. The first step is to remove the effect of the sample mean.
Denote $m_{N,j}=\frac{1}{N}{\bf z}_j^T{\bf M}_{N,j}{\bf z}_j$ and ${\bf M}_{N,j}\!\!=\!\left(\!\frac{1-\rho_n}{1-(1-\rho_n)c_N}\frac{1}{n} \!\sum_{t\neq j}{\bf z}_{t}{\bf z}_{t}^T\!+({\gamma}(\rho_n)\!+\ell)\rho_n{\bf I}_N\!\right)^{-1}$.
We have in particular:
\begin{prop} \label{prop: 1}
As $N,n\rightarrow\infty$,
\begin{align}\label{eq:prop1}
\max_{1\leq j\leq n}\left|\widetilde{m}_{N,j}-m_{N,j}\right|\stackrel{a.s.}\longrightarrow0.
\end{align}
\end{prop}

\emph{Proof:} See Appendix \ref{prof: prop1}.
\begin{remark}
In Proposition \ref{prop: 1}, Assumption \ref{assump}-\ref{assump:2} is necessary; that is,  i.i.d. $\tau_1,...,\tau_n\geq\xi$ a.s. for some $\xi>0$ and $E[\tau_1]<\infty$. It guarantees that for $t=1,...,n$, the norm of $\tilde{\bf z}_t$ does not go off to infinity, recalling that $\tilde{\bf z}_{t}={\bf z}_t-\frac{1}{n}{\bf Z}_N\sqrt{\frac{\pmb \tau}{\tau_t}}$.
\end{remark}

By Proposition \ref{prop: 1},
we have $\left|\widetilde{m}_{N,n}-m_{N,n}\right|\stackrel{\rm a.s.}\longrightarrow0$. This allows us to follow the proof in \cite{couillet2014large}, which deals with data with mean zero.

To proceed, assume first $\rho_0\neq 1$. From the proof of Theorem $1$ in \cite{couillet2014large},
\begin{align} \nonumber
&m_{N,n}\stackrel{\rm a.s.}\longrightarrow \\ \label{eq:eplus}
&\frac{1-(1-\rho_0)c}{1-\rho_0}\delta\left(-({\gamma}(\rho_0)+\ell)\rho_0\frac{1-(1-\rho_0)c}{1-\rho_0}\right)\triangleq m^+,
\end{align}
where, for $x<0$, $\delta(x)$ is the unique positive solution to
\begin{align} \nonumber
\delta(x)=\int\frac{t}{-x+\frac{t}{1+c\delta(x)}}\nu(dt).
\end{align}
Together with $\left|\widetilde{m}_{N,n}-m_{N,n}\right|\stackrel{\rm a.s.}\longrightarrow0$, we have
\begin{align} \label{eq:maxconv}
&\left|\widetilde{m}_{N,n}-m^+\right|\stackrel{\rm a.s.}\longrightarrow0.
\end{align}
It was demonstrated in \cite{couillet2014large} that $m^+<1$. But this is in contradiction with $\widetilde{m}_{N,n}\geq1$.

Now assume $\rho_0=1$. According to \cite{couillet2014large},
\begin{align} \nonumber
m_{N,n}\stackrel{\rm a.s.}\longrightarrow\frac{1}{1+\ell}<1.
\end{align}
Then
\begin{align} \nonumber
\left|\widetilde{m}_{N,n}-\frac{1}{1+\ell}\right|\stackrel{\rm a.s.}\longrightarrow0,
\end{align}
but $\frac{1}{1+\ell}<1$, again raising a contradiction with $\widetilde{m}_{N,n}\geq1$.

Hence, for all large $n$, there is no converging subsequence of $\rho_n$ (and thus no subsequence of $\rho_n$) for which $\hat{d}_{{n}}(\rho_n)>\gamma(\rho_n)+\ell$ infinitely often. Therefore $\hat{d}_{{n}}(\rho)\leq\gamma(\rho)+\ell$ for all large $n$ a.s., uniformly on $\rho\in{\mathcal{R}}_\varepsilon$.

The same reasoning holds for $\hat{d}_1(\rho)$, which can be proved greater than $\gamma(\rho)-\ell$ for all large $n$ uniformly on $\rho\in\mathcal{R}_{\varepsilon}$. Following the same arguments in \cite{couillet2014large}, since $\ell>0$ is arbitrary, from the ordering of the $\hat{d}_t(\rho)$, we have proved that $\sup_{\rho\in{\mathcal{R}}_\varepsilon}\max_{1\leq t\leq n}\left|\hat{d}_t(\rho)-{\gamma}(\rho)\right|\stackrel{\rm a.s.}\longrightarrow0$.~~~~~~~~~$\blacksquare$

The following three lemmas, Lemma \ref{lemma:FR}, \ref{lemma:FR_2} and \ref{lemma:FR2} show that functionals of Tyler's estimator asymptotically perform similar to functionals of $\frac{1}{n}\sum_{t=1}^n{\bf z}_{t}{\bf z}_{t}^T$ or $\frac{1}{n}\sum_{t=1}^n{\bf y}_{t}{\bf y}_{t}^T$. They are used as an intermediate step for the development of the asymptotic deterministic equivalent of the risk function. Using existing results in \cite{Fran2012}, quoted as Lemma \ref{lemma:Fran} in this paper, we can then obtain our main theorems.

For notational convenience, we denote $k=k(\rho)\triangleq\frac{1-\rho}{1-(1-\rho)c}$. Also recall that $\gamma$ is the unique positive solution to $1=\int\frac{t}{\gamma\rho+(1-\rho)t}\nu(dt)$. Assuming ${\bf A}_N\in\mathbb{R}^{N\times N}$ is a deterministic symmetric nonnegative definite matrix, for some $\eta>0$, define $\mathcal{D}=
\begin{cases}
[0,\infty)& \text{if}~\liminf_N\lambda_1({\bf A}_N)>0\\
[\eta,\infty)& \text{otherwise}
\end{cases}$,
and further define that, for \mbox{$\rho\in\mathcal{R}_\varepsilon$ and $w\in\mathcal{D}$,}\
\begin{align} \nonumber
\widetilde{\bf R}_N&=\left({\bf A}_N+(1-\rho)\frac{1}{n}\sum_{t=1}^n\frac{\tilde{\bf x}_{t}\tilde{\bf x}_{t}^{T}}{\frac{1}{N}\tilde{\bf x}_{t}^T\hat{\bf C}_{\rm ST}^{-1}\tilde{\bf x}_{t}}+w{\bf I}_N\right)^{-1} \\ \nonumber
\widetilde{\bf S}_N&=\left({\bf A}_N+\frac{k}{\gamma}\frac{1}{n}\sum_{t=1}^n\tilde{\bf z}_{t}\tilde{\bf z}_{t}^T+w{\bf I}_N\right)^{-1} \\ \nonumber
{\bf S}_N&=\left({\bf A}_N+\frac{k}{\gamma}\frac{1}{n}\sum_{t=1}^n{\bf z}_{t}{\bf z}_{t}^T+w{\bf I}_N\right)^{-1}.
\end{align}
Then we introduce the following lemma.

\begin{lemma} \label{lemma:FR}
Assume ${\bf a}_N\in\mathbb{R}^N$ is a deterministic vector with $\lim\sup_N\|{\bf a}_N\|^2<\infty$.
Under the settings of Theorem \ref{thrm:asymp}, as $N,n\rightarrow\infty$,
\begin{align}\label{eq:asymptrace_sigma}
\sup_{\rho\in\mathcal{R}_\varepsilon,w\in\mathcal{D}}\left|{\bf a}_N^T\widetilde{\bf R}_N{\bf a}_N-{\bf a}_N^T{\bf S}_N{\bf a}_N\right|\stackrel{\rm a.s.}\longrightarrow0.
\end{align}
\end{lemma}
\emph{Proof:}
 Define
 \begin{align} \nonumber
 \hat{\bf B}_N(\rho)&=\frac{k}{{\gamma(\rho)}}\frac{1}{n}\sum_{t=1}^n\tilde{\bf z}_{t}\tilde{\bf z}_{t}^T, \\ \nonumber
\hat{\bf D}_N(\rho)&=(1-\rho)\frac{1}{n}\sum_{t=1}^n\frac{\tilde{\bf x}_{t}\tilde{\bf x}_{t}^T}{\frac{1}{N}\tilde{\bf x}_{t}^T\hat{\bf C}_{\rm ST}^{-1}\tilde{\bf x}_{t}}  \\ \nonumber
&\stackrel{(a)}=\frac{1-\rho}{1-(1-\rho)c_N}\frac{1}{n}\sum_{t=1}^n\frac{\tilde{\bf z}_{t}\tilde{\bf z}_{t}^T}{\hat{d}_t(\rho)}
\end{align}
where $(a)$ uses the identity (\ref{eq:change}).
Denote
\begin{align} \nonumber
\Delta&\triangleq{\bf a}_N^T\widetilde{\bf R}_N{\bf a}_N-{\bf a}_N^T\widetilde{\bf S}_N{\bf a}_N \\ \nonumber
&\stackrel{(a)}={\bf a}_N^T\widetilde{\bf R}_N\left(\hat{\bf B}_N(\rho)-\hat{\bf D}_N(\rho)\right)\widetilde{\bf S}_N{\bf a}_N
\end{align}
where $(a)$  uses the identity that ${\bf U}^{-1}-{\bf V}^{-1}={\bf U}^{-1}({\bf V}-{\bf U}){\bf V}^{-1}$ for invertible ${\bf U},{\bf V}$ matrices.
We first prove that as $N,n\rightarrow\infty$, $\sup_{\rho\in\mathcal{R}_\varepsilon,w\in\mathcal{D}}\left|\Delta\right|\stackrel{\rm a.s.}\longrightarrow0$.

As $N, n\rightarrow\infty$, using the definition of $k$,
\begin{align} \nonumber
&\sup_{\rho\in{\mathcal{R}}_\varepsilon}\left\|\hat{\bf D}_N(\rho)-\hat{\bf B}_N(\rho)\right\| \\ \label{eq:diffspec}
&\leq\left\|\frac{1}{n}\sum_{t=1}^n\tilde{\bf z}_{t}\tilde{\bf z}_{t}^T\right\|\sup_{\rho\in{\mathcal{R}}_\varepsilon}\max_{1\leq t\leq n}\frac{1-\rho}{1-(1-\rho)c}\left|\frac{\hat{d}_t(\rho)-\gamma(\rho)}{\gamma(\rho)\hat{d}_t(\rho)}\right|.
\end{align}
We will show that the RHS of (\ref{eq:diffspec}) goes to $0$ a.s. Recalling Lemma \ref{lemma:dgamma}, this follows upon showing that $\limsup_n\left\|\frac{1}{n}\sum_{t=1}^n\tilde{\bf z}_{t}\tilde{\bf z}_{t}^T\right\|<\infty$ a.s. To this end, recall that $\tilde{\bf z}_{t}={\bf z}_t-\frac{1}{n}{\bf Z}_N\sqrt{\frac{\pmb \tau}{\tau_t}}$. Then
\begin{align} \nonumber
&\frac{1}{n}\sum_{t=1}^n\tilde{\bf z}_{t}\tilde{\bf z}_{t}^T \\ \nonumber
&=\!\frac{1}{n}\!\sum_{t=1}^n\!{\bf z}_{t}{\bf z}_{t}^T\!\!-\!\frac{1}{n}\!\sum_{t=1}^n{\bf z}_t\!\left(\!\frac{1}{n}{\bf Z}_N\frac{\sqrt{\pmb\tau}}{\sqrt{\tau_t}}\right)^T\!\!\!\!-\!\frac{1}{n}\!\sum_{t=1}^n\!\left(\frac{1}{n}{\bf Z}_N\frac{\sqrt{\pmb\tau}}{\sqrt{\tau_t}}\right){\bf z}_t^T  \\  \label{eq:specdisect}
&+\frac{1}{n}\sum_{t=1}^n\left(\frac{1}{n}{\bf Z}_N\frac{\sqrt{\pmb\tau}}{\sqrt{\tau_t}}\right)\left(\frac{1}{n}{\bf Z}_N\frac{\sqrt{\pmb\tau}}{\sqrt{\tau_t}}\right)^T.
\end{align}
We will show that the spectral norm of each term on the RHS of (\ref{eq:specdisect}) is bounded for all large $n$ a.s.

First, from Assumption \ref{assump}-\ref{assump:3}. and \cite{Bai&Silverstein1998}, we have $\limsup_n\left\|\frac{1}{n}\sum_{t=1}^n{\bf z}_{t}{\bf z}_{t}^T\right\|<\infty$ a.s. Next, for the second and the third terms on the RHS of (\ref{eq:specdisect}),
\begin{align} \nonumber
&\left\|\frac{1}{n}\sum_{t=1}^n{\bf z}_t\left(\frac{1}{n}{\bf Z}_N\frac{\sqrt{\pmb\tau}}{\sqrt{\tau_t}}\right)^T\right\|=\left\|\frac{1}{n}\sum_{t=1}^n\left(\frac{1}{n}{\bf Z}_N\frac{\sqrt{\pmb\tau}}{\sqrt{\tau_t}}\right){\bf z}_t^T\right\| \\ \nonumber
&=\left(\frac{1}{n}\sum_{t=1}^n\frac{{\bf y}_t}{\sqrt{\tau_t}}\right)^T{\bf C}_N\left(\frac{1}{n}\sum_{t=1}^n{\bf y}_t\sqrt{\tau_t}\right) \\ \nonumber 
&\leq\left\|{\bf C}_N\right\|\left(\frac{1}{n}\sum_{t=1}^n\frac{{\bf y}_t}{\sqrt{\tau_t}}\right)^T\left(\frac{1}{n}\sum_{t=1}^n{\bf y}_t\sqrt{\tau_t}\right).
\end{align}
By the law of large numbers, as $N, n\rightarrow\infty$,
\begin{align} \nonumber
\left|\left(\frac{1}{n}\sum_{t=1}^n\frac{{\bf y}_t}{\sqrt{\tau_t}}\right)^T\left(\frac{1}{n}\sum_{t=1}^n{\bf y}_t\sqrt{\tau_t}\right)-c\right|\stackrel{\rm a.s.}\longrightarrow0.
\end{align}
According to Assumption \ref{assump}-\ref{assump:1} and Assumption \ref{assump}-\ref{assump:3}, we can see that $\limsup_n\left\|\frac{1}{n}\sum_{t=1}^n{\bf z}_t\left(\frac{1}{n}{\bf Z}_N\frac{\sqrt{\pmb\tau}}{\sqrt{\tau_t}}\right)^T\right\|=\limsup_n\left\|\frac{1}{n}\sum_{t=1}^n\left(\frac{1}{n}{\bf Z}_N\frac{\sqrt{\pmb\tau}}{\sqrt{\tau_t}}\right){\bf z}_t^T\right\|\leq c\left\|{\bf C}_N\right\|<\infty$\,\;a.s.

For the fourth term, with Assumption \ref{assump}-\ref{assump:2}, we have
 \begin{align} \nonumber
&\limsup_n\left\|\frac{1}{n}\sum_{t=1}^n\left(\frac{1}{n}{\bf Z}_N\frac{\sqrt{\pmb\tau}}{\sqrt{\tau_t}}\right)\left(\frac{1}{n}{\bf Z}_N\frac{\sqrt{\pmb\tau}}{\sqrt{\tau_t}}\right)^T\right\| \\ \nonumber
&\leq\limsup_n\!\left\{\left\|\frac{1}{n}{\bf Z}_N^T{\bf Z}_N\right\|\!\left(\frac{1}{n}\sum_{t=1}^n\tau_t\right)\!\left(\frac{1}{n}\sum_{t=1}^n\frac{1}{\tau_t}\right)\!\right\}\!<\!\infty ~\text{a.s.}
\end{align}
Therefore, $\limsup_n\left\|\frac{1}{n}\sum_{t=1}^n\tilde{\bf z}_{t}\tilde{\bf z}_{t}^T\right\|<\infty$ a.s. Together with Lemma \ref{lemma:dgamma}, from (\ref{eq:diffspec}), we have
\begin{align} \label{eq:etaspectrum}
\sup_{\rho\in{\mathcal{R}}_\varepsilon}\left\|\hat{\bf B}_N(\rho)-\hat{\bf D}_N(\rho)\right\|\stackrel{\rm a.s.}\longrightarrow0.
\end{align}

Note that $w\in\mathcal{D}$ ensures $\limsup_N\sup_{\rho\in\mathcal{R}_\varepsilon,w\in\mathcal{D}}\left\|\widetilde{\bf R}_N\right\|<\infty$ and $\limsup_N\sup_{\rho\in\mathcal{R}_\varepsilon,w\in\mathcal{D}}\left\|\widetilde{\bf S}_N\right\|<\infty$.

Together with (\ref{eq:etaspectrum}) and $\left\|{\bf a}_N\right\|^2<\infty$, we have
\begin{align}
\nonumber
&\sup_{\rho\in\mathcal{R}_\varepsilon,w\in\mathcal{D}}|\Delta|\leq\|{\bf a}_N\|^2 \sup_{\rho\in\mathcal{R}_\varepsilon,w\in\mathcal{D}}\left\|\widetilde{\bf R}_N\right\|\sup_{\rho\in\mathcal{R}_\varepsilon,w\in\mathcal{D}}\left\|\widetilde{\bf S}_N\right\|\\ \nonumber
&\times\!\sup_{\rho\in\mathcal{R}_\varepsilon}\left\|\frac{k}{{\gamma}}\frac{1}{n}\sum_{t=1}^n\tilde{\bf z}_{t}\tilde{\bf z}_{t}^T\!-(1-\rho)\frac{1}{n}\sum_{t=1}^n\frac{\tilde{\bf x}_{t}\tilde{\bf x}_{t}^T}{\frac{1}{N}\tilde{\bf x}_{t}^T\hat{\bf C}_{\rm ST}^{-1}\tilde{\bf x}_{t}}\right\|\stackrel{\rm a.s.}\longrightarrow0.
\end{align}
\normalsize
Following the same reasoning as that of Proposition \ref{prop: 1}, we have
\begin{align} \nonumber
&\sup_{\rho\in\mathcal{R}_\varepsilon,w\in\mathcal{D}}\left|{\bf a}_N^T\widetilde{\bf S}_N{\bf a}_N-{\bf a}_N^T{\bf S}_N{\bf a}_N\right|\stackrel{\rm a.s.}\longrightarrow0.
\end{align}
Together with $\sup_{\rho\in\mathcal{R}_\varepsilon,w\in\mathcal{D}}\left|\Delta\right|\stackrel{\rm a.s.}\longrightarrow0$,  we obtain (\ref{eq:asymptrace_sigma}). $~~~~\blacksquare$

Define ${\bf W}_N=\left(\!{\bf A}_N+\frac{k}{\gamma}\frac{1}{n}\sum_{t=1}^n{\bf y}_{t}{\bf y}_{t}^T+w{\bf I}_N\right)^{-1}$ and $\widetilde{\bf W}_N=\left({\bf A}_N+(1-\rho)\frac{1}{n}\sum_{t=1}^n\frac{\tilde{\bf y}_{t}\tilde{\bf y}_{t}^{T}}{\frac{1}{N}\tilde{\bf y}_{t}^T{\bf C}_N^{1/2}\hat{\bf C}_{\rm ST}^{-1}{\bf C}_N^{1/2}\tilde{\bf y}_{t}}+w{\bf I}_N\!\right)^{-1}$, where $\tilde{\bf y}_{t}={\bf y}_{t}-\frac{1}{n}\sum_{i=1}^n{\bf y}_{i}$. We introduce the following lemma.
\begin{lemma} \label{lemma:FR_2}
Under the settings of Lemma \ref{lemma:FR}, as $N,n\rightarrow\infty$,
\begin{align} \label{eq:asymptrace_nosigma}
\sup_{\rho\in\mathcal{R}_\varepsilon,w\in\mathcal{D}}\left|{\bf a}_N^T\widetilde{\bf W}_N{\bf a}_N-{\bf a}_N^T{\bf W}_N{\bf a}_N\right|\stackrel{\rm a.s.}\longrightarrow0.
\end{align}
\end{lemma}
\emph{Proof:} The derivation is similar to that of (\ref{eq:asymptrace_sigma}).

 \begin{lemma} \label{lemma:FR2}
 Under the settings of Lemma \ref{lemma:FR} and assuming ${\bf A}_N={\bf 0}$, as $N,n\rightarrow\infty$,
 \begin{align} \nonumber
\sup_{\rho\in\mathcal{R}_\varepsilon,w\in[\eta,\infty)}\left|{\bf a}_N^T(1-\rho)\frac{1}{n}\sum_{i=1}^n\frac{\tilde{\bf x}_{i}\tilde{\bf x}_{i}^{T}}{\frac{1}{N}\tilde{\bf x}_{i}^T\hat{\bf C}_{\rm ST}^{-1}\tilde{\bf x}_{i}}\widetilde{\bf R}_N^2{\bf a}_N \right.\\ \label{eq:lemmaFR2}
\left.-{\bf a}_N^T\frac{k}{\gamma}\frac{1}{n}\sum_{i=1}^n{\bf z}_{i}{\bf z}_{i}^T{\bf S}_N^2{\bf a}_N\right|\stackrel{\rm a.s.}\longrightarrow0.
\end{align}
\end{lemma}
\emph{Proof:}
We first notice that
\begin{align}\nonumber
&{\bf a}_N^T(1-\rho)\frac{1}{n}\sum_{i=1}^n\frac{\tilde{\bf x}_{i}\tilde{\bf x}_{i}^{T}}{\frac{1}{N}\tilde{\bf x}_{i}^T\hat{\bf C}_{\rm ST}^{-1}\tilde{\bf x}_{i}}\widetilde{\bf R}_N^{2}{\bf a}_N\\ \nonumber
&=-\frac{\rm d}{{\rm d}w}\left[{\bf a}_N^T(1-\rho)\frac{1}{n}\sum_{i=1}^n\frac{\tilde{\bf x}_{i}\tilde{\bf x}_{i}^{T}}{\frac{1}{N}\tilde{\bf x}_{i}^T\hat{\bf C}_{\rm ST}^{-1}\tilde{\bf x}_{i}}\widetilde{\bf R}_N{\bf a}_N\right]\\ \nonumber
&=-\frac{\rm d}{{\rm d}w}\left({\bf a}_N^T{\bf a}_N\!-w{\bf a}_N^T\widetilde{\bf R}_N{\bf a}_N\right)\\ \nonumber
&={\bf a}_N^T\widetilde{\bf R}_N{\bf a}_N+w\frac{\rm d}{{\rm d}w}\left({\bf a}_N^T\widetilde{\bf R}_N{\bf a}_N\right).
\end{align}
Following similar steps, we also have
\begin{align}\nonumber
{\bf a}_N^T\frac{k}{\gamma}\frac{1}{n}\sum_{i=1}^n{\bf z}_{i}{\bf z}_{i}^T{\bf S}_N^2{\bf a}_N={\bf a}_N^T{\bf S}_N{\bf a}_N+w\frac{\rm d}{{\rm d}w}\left({\bf a}_N^T{\bf S}_N{\bf a}_N\right).
\end{align}
The almost sure convergence (\ref{eq:asymptrace_sigma}) in Lemma \ref{lemma:FR} when extended to $w\in\mathbb{C}$ is uniform on any bounded region of $\left(\mathbb{C}\/- \mathbb{R}\right)\cup\mathcal{D}$, and the functionals of $w$ in (\ref{eq:asymptrace_sigma}) are analytic. Thus, by the Weierstrass convergence theorem \cite{Ahlforscomplexanalysis}, the following holds:
\begin{align}\nonumber
\sup_{\rho\in\mathcal{R}_\varepsilon,w\in\mathcal{D}}\left|\frac{\rm d}{{\rm d}w}\left({\bf a}_N^T\widetilde{\bf R}_N{\bf a}_N\right)-\frac{\rm d}{{\rm d}w}\left({\bf a}_N^T{\bf S}_N{\bf a}_N\right)\right|\stackrel{\rm a.s.}\longrightarrow0.
\end{align}
Together with Lemma \ref{lemma:FR}, we obtain (\ref{eq:lemmaFR2}).$~~~~~~~~~~~~~~~~~~~~\blacksquare$


\begin{lemma}\cite[Appendix I-B]{Fran2012} \label{lemma:Fran}
Under the settings of Lemma~\ref{lemma:FR}, as $N,n\rightarrow\infty$,
\begin{align} \label{eq:Fran}
\sup_{\rho\in\mathcal{R}_\varepsilon,w\in\mathcal{D}}\left|{\bf a}_N^T{\bf S}_N{\bf a}_N-{\bf a}_N^T{\bf T}_N{\bf a}_N\right|\stackrel{\rm a.s.}\longrightarrow0
\end{align}
where ${\bf T}_N=\left({\bf A}_N+\frac{k}{(\gamma+e_N(w)k)}{\bf C}_N+w{\bf I}_N\right)^{-1}$ and, for each $w\in\mathcal{D}$, $e_N(w)$ is the unique positive solution to the following equation:
\begin{align}
\nonumber e_N(w)=\frac{1}{n}{\rm tr}\left[{\bf C}_N\!\left(\!{\bf A}_N+\frac{k}{(\gamma+e_N(w)k)}{\bf C}_N+w{\bf I}_N\right)^{-1}\right].
\end{align}
Moreover, when ${\bf A}_N={\bf 0}$, we have
\small
\begin{align} \nonumber
&\sup_{\rho\in\mathcal{R}_\varepsilon,w\in[\eta,\infty)}\left|{\bf a}_N^T\frac{k}{\gamma}\frac{1}{n}\sum_{t=1}^n{\bf z}_{t}{\bf z}_{t}^T{\bf S}_N^2{\bf a}_N\right. \\ \nonumber
&\left.-\frac{k\gamma}{(\gamma+e_N(w) k)^2}\frac{{\bf a}_N^T{\bf C}_N{\bf T}_N^2{\bf a}_N}{1-\frac{k^2}{(\gamma+e_N(w)k)^2}\dfrac{1}{n}{\rm tr}\left[{\bf C}_N^2{\bf T}_N^2\right]}\right|\stackrel{\rm a.s.}\longrightarrow0.~~~~\blacksquare
\end{align}
\normalsize
\end{lemma}



\section{Proof of Theorem \ref{thrm:asymp}} \label{appx_prfth1}
First consider the (re-scaled) realized portfolio risk:
\begin{align} \label{eq:SFPE rescaled risk}
N\sigma^2(\hat{\bf h}_{\rm ST})=\frac{\frac{1}{N}{\bf 1}_N^T\hat{\bf C}_{\rm ST}^{-1}{\bf C}_N\hat{\bf C}_{\rm ST}^{-1}{\bf 1}_N}{(\frac{1}{N}{\bf 1}_N^T\hat{\bf C}_{\rm ST}^{-1}{\bf 1}_N)^2}.
\end{align}

For the denominator, Lemma \ref{lemma:FR} and Lemma \ref{lemma:Fran} imply
\small
\begin{align} \nonumber 
&\sup_{\rho\in\mathcal{R}_\varepsilon}\left|\dfrac{1}{N}{\bf 1}_N^T\hat{\bf C}^{-1}_{\rm ST}{\bf 1}_N-\dfrac{1}{N}{\bf 1}_N^T\left(\frac{1-\rho}{\gamma}{\bf C}_N+\rho{\bf I}_N\right)^{-1}{\bf 1}_N\right|\stackrel{\rm a.s.}\longrightarrow0.
\end{align}
\normalsize
Note that in this case, ${\bf A}_N={\bf 0}$, ${\bf a}_N=\frac{1}{\sqrt{N}}{\bf 1}_N$ and $w=\rho$, which leads to $\left|e_N(\rho)-c\gamma\right|\stackrel{\rm a.s.}\longrightarrow0$ when $N, n\rightarrow\infty$. The derivation is based on Assumption \ref{assump}-\ref{assump:3} and the definition of $\gamma$ in (\ref{def:gamma}).

For the numerator, we rewrite it as
\begin{align}
\nonumber&\frac{1}{N}{\bf 1}_N^T\hat{\bf C}_{\rm ST}^{-1}{\bf C}_N\hat{\bf C}_{\rm ST}^{-1}{\bf 1}_N \\ \nonumber
&=\frac{1}{N}{\bf 1}_N^T{\bf C}_N^{-1/2}({\bf C}_N^{-1/2}\hat{\bf C}_{\rm ST}{\bf C}_N^{-1/2})^{-2}{\bf C}_N^{-1/2}{\bf 1}_N,
\end{align}
which, upon substituting the RHS of (\ref{eq:SFPE}) for $\hat{\bf C}_{\rm ST}$ and setting ${\bf A}_N=\rho{\bf C}_N^{-1}$ in $\widetilde{\bf W}_N$, yields
\begin{align}
\nonumber&\frac{1}{N}{\bf 1}_N^T\hat{\bf C}_{\rm ST}^{-1}{\bf C}_N\hat{\bf C}_{\rm ST}^{-1}{\bf 1}_N
\\ \nonumber
&=\frac{1}{N}{\bf 1}_N^T{\bf C}_N^{-1/2}\!\!\left(\!(1\!-\!\rho)\frac{1}{n}\!\sum_{t=1}^n\!\frac{{\bf C}_N^{-1/2}\tilde{\bf x}_{t}\tilde{\bf x}_{t}^T{\bf C}_N^{-1/2}}{\frac{1}{N}\tilde{\bf x}_{t}^T\hat{\bf C}_{\rm ST}^{-1}\tilde{\bf x}_{t}}\!+\!\rho{\bf C}_N^{-1}\!\right)^{-2} \\ \nonumber
&~~~\times{\bf C}_N^{-1/2}{\bf 1}_N \\ \nonumber
&\left.=-\frac{\rm d}{{\rm d}w}\left[\frac{1}{N}{\bf 1}_N^T{\bf C}_N^{-1/2}\widetilde{\bf W}_N{\bf C}_N^{-1/2}{\bf 1}_N\right]\right|_{w=0}.
\end{align}

Setting ${\bf A}_N=\rho{\bf C}_N^{-1}$ and ${\bf a}_N^T=\frac{1}{\sqrt{N}}{\bf C}_N^{-1/2}{\bf 1}_N$ in (\ref{eq:Fran}), as well as ${\bf A}_N=\rho{\bf C}_N^{-1}$ in ${\bf W}_N$, yields
\begin{align} \nonumber
\sup_{\rho\in\mathcal{R}_\varepsilon,w\in[0,\infty)}\left|\frac{1}{N}{\bf 1}_N^T{\bf C}_N^{-1/2}{\bf W}_N{\bf C}_N^{-1/2}{\bf 1}_N\right. \\ \label{eq:Fran_extention}
\left.-\frac{1}{N}{\bf 1}_N^T{\bf C}_N^{-1/2}{\bf J}_N{\bf C}_N^{-1/2}{\bf 1}_N\right|\stackrel{\rm a.s.}\longrightarrow0
\end{align}
where ${\bf J}_N=\left(\rho{\bf C}_N^{-1}+\left(\frac{k}{(\gamma+e_N(w)k)}+w\right){\bf I}_N\right)^{-1}$ and for each $w\in[0,\infty)$, $\tilde{e}_N(w)$ is the unique positive solution to the following equation:
\begin{align}
\nonumber \tilde{e}_N(w)&=\frac{1}{n}{\rm tr}\left[\left(\rho{\bf C}_N^{-1}+\left(\frac{k}{\gamma+\tilde{e}_N(w)k}+w\right){\bf I}_N\right)^{-1}\right].
\end{align}


Lemma \ref{lemma:FR_2} and the convergence (\ref{eq:Fran_extention}) imply
\begin{align}
\nonumber&\sup_{\rho\in\mathcal{R}_\varepsilon,w\in[0,\infty)}\left|\frac{1}{N}{\bf 1}_N^T{\bf C}_N^{-1/2}\widetilde{\bf W}_N{\bf C}_N^{-1/2}{\bf 1}_N\right. \\ \nonumber
&\left.~~~~~~~~~~~~~~~~~-\frac{1}{N}{\bf 1}_N^T{\bf C}_N^{-1/2}{\bf J}_N{\bf C}_N^{-1/2}{\bf 1}_N\right|\stackrel{\rm a.s.}\longrightarrow0.
\end{align}
Following the same reasoning as for the proof of Lemma \ref{lemma:FR2}, the convergence of the derivatives holds such that at $w=0$ by the Weierstrass convergence theorem,
\begin{align}
\nonumber&\sup_{\rho\in\mathcal{R}_\varepsilon}\left|\frac{\rm d}{{\rm d}w}\left(\frac{1}{N}{\bf 1}_N^T{\bf C}_N^{-1/2}\widetilde{\bf W}_N{\bf C}_N^{-1/2}{\bf 1}_N\right)\right|_{w=0}\\ \nonumber
&\left.~~~~~~~\left.-\frac{\rm d}{{\rm d}w}\left(\frac{1}{N}{\bf 1}_N^T{\bf C}_N^{-1/2}{\bf J}_N{\bf C}_N^{-1/2}{\bf 1}_N\right)\right|_{w=0}
\right|\stackrel{\rm a.s.}\longrightarrow0.
\end{align}

With Eq. (\ref{eq:derivative}) on the top of the next page
\begin{figure*}
\begin{align} \label{eq:derivative}
&\left.\frac{\rm d}{{\rm d}w}\left[\frac{1}{N}{\bf 1}_N^T{\bf C}_N^{-1/2}{\bf J}_N{\bf C}_N^{-1/2}{\bf 1}_N\right]\right|_{w=0}=\frac{-\frac{1}{N}{\bf 1}_N^T{\bf C}_N^{1/2}\left(\frac{k}{\gamma+\tilde{e}_N(0)k}{\bf C}_N+\rho{\bf I}_N\right)^{-2}{\bf C}_N^{1/2}{\bf 1}_N}{1-\frac{k^2}{(\gamma+\tilde{e}_N(0)k)^2}\dfrac{1}{n}{\rm tr}\left[{\bf C}_N^2\left(\left(\frac{k}{\gamma+\tilde{e}_N(0)k}\right){\bf C}_N+\rho{\bf I}_N\right)^{-2}\right]}.
\end{align}
\hrule
\end{figure*}
and $\left|\tilde{e}_N(0)-c\gamma\right|\stackrel{\rm a.s.}\longrightarrow0$ when $N, n\rightarrow\infty$, we have
\begin{align}
\nonumber&\sup_{\rho\in\mathcal{R}_\varepsilon}\left|\frac{1}{N}{\bf 1}_N ^T\hat{\bf C}_{\rm ST}^{-1}{\bf C}_N\hat{\bf C}_{\rm ST}^{-1}{\bf 1}_N-\right. \frac{\gamma^2}{\gamma^2-\beta(1-\rho)^2}\frac{1}{N}{\bf 1}_N^T\\ \nonumber
&\left.\times\left(\frac{1-\rho}{\gamma}{\bf C}_N+\rho{\bf I}_N\right)^{-1}\!\!\!\!\!{\bf C}_N\left(\frac{1-\rho}{\gamma}{\bf C}_N+\rho{\bf I}_N\right)^{-1}{\bf 1}_N\right| \\ \label{eq:asympNMR}
&\stackrel{\rm a.s.}\longrightarrow0.
\end{align}
Equipped with the asymptotic equivalences of the denominator and numerator of (\ref{eq:SFPE rescaled risk}), we prove Theorem \ref{thrm:asymp}.

\section{Proof of Lemma \ref{lemma:gamma_est}} \label{appx_prflemma1}
First notice that
\begin{align} \nonumber
{\hat{\gamma}_{\rm sc}}&=\dfrac{1}{1-(1-\rho)c_N}\frac{1}{N}\dfrac{1}{n}\sum_{t=1}^n\frac{\tilde{\bf x}_{t}^T\hat{\bf C}_{\rm ST}^{-1}(\rho)\tilde{\bf x}_{t}}{\frac{1}{N}\|\tilde{\bf x}_{t}\|^2} \\ \nonumber
&=\dfrac{1}{1-(1-\rho)c_N}\frac{1}{N}\dfrac{1}{n}\sum_{t=1}^n\frac{\tilde{\bf z}_{t}^T\hat{\bf C}_{\rm ST}^{-1}(\rho)\tilde{\bf z}_{t}}{\frac{1}{N}\|\tilde{\bf z}_{t}\|^2}.
\end{align}
It has been shown in Lemma \ref{lemma:dgamma} that $\sup_{\rho\in{\mathcal{R}}_\varepsilon}\max_{1\leq t\leq n}\left|\hat{d}_t(\rho)-{\gamma}(\rho)\right|\stackrel{\rm a.s.}\longrightarrow0$, where $\hat{d}_t(\rho)=\frac{1}{1-(1-\rho)c_N}\frac{1}{N}\tilde{\bf z}_{t}^T\hat{\bf C}_{\rm ST}^{-1}(\rho)\tilde{\bf z}_{t}$.
Therefore, to prove the convergence (\ref{eq:estgamma}), it is left to show that $\frac{1}{N}\|\tilde{\bf z}_{t}\|^2\stackrel{\rm a.s.}\longrightarrow\kappa$.

We start by writing
\begin{align} \nonumber
\frac{1}{N}\|\tilde{\bf z}_{t}\|^2=\frac{1}{N}\left({\bf z}_{t}^T{\bf z}_{t}-\frac{1}{n}{\bf z}_{t}^T{\bf Z}_N\frac{\sqrt{\pmb\tau}}{\sqrt{\tau_{t}}}-\frac{1}{n}\frac{\sqrt{\pmb\tau}^T}{\sqrt{\tau_{t}}}{\bf Z}_N^T{\bf z}_{t}\right. \\  \label{eq:z_tilde}
\left.+\frac{1}{n^2}\frac{\sqrt{\pmb\tau}^T}{\sqrt{\tau_{t}}}{\bf Z}_N^T{\bf Z}_N\frac{\sqrt{\pmb\tau}}{\sqrt{\tau_{t}}}\right).
\end{align}
Since the second and the third term on the RHS of (\ref{eq:z_tilde}) are the same, we analyze the second term only. It can be rewritten as
\begin{align} \nonumber
\frac{1}{N}\frac{1}{n}{\bf z}_{t}^T{\bf Z}_N\frac{\sqrt{\pmb\tau}}{\sqrt{\tau_{t}}}=\frac{1}{N}\frac{1}{n}{\bf z}_{t}^T{\bf z}_{t}+\frac{1}{N}\frac{1}{n}{\bf z}_{t}^T{\bf Z}_N^{(t)}\frac{\sqrt{{\pmb\tau}^{(t)}}}{\sqrt{\tau_{t}}}
\end{align}
where ${\bf Z}_N^{(t)}$ is the matrix with the $t$th column removed from ${\bf Z}_N$ and $\sqrt{{\pmb\tau}^{(t)}}$ is the vector with the $t$th entry removed from $\sqrt{\pmb\tau}$. Since ${\bf z}_t$ is independent of $\frac{1}{n}{\bf Z}_N^{(t)}\frac{\sqrt{{\pmb\tau}^{(t)}}}{\sqrt{\tau_{t}}}$, we have $\frac{1}{n}{\bf z}_{t}^T{\bf Z}_N^{(t)}\frac{\sqrt{{\pmb\tau}^{(t)}}}{\sqrt{\tau_{t}}}\stackrel{\rm a.s.}\longrightarrow0$. Together with $\frac{1}{n}{\bf z}_{t}^T{\bf z}_{t}=O(1)$ a.s., we have $\frac{1}{N}\frac{1}{n}{\bf z}_{t}^T{\bf Z}_N\frac{\sqrt{\pmb\tau}}{\sqrt{\tau_{t}}}\stackrel{\rm a.s.}\longrightarrow0$.

For the last term in (\ref{eq:z_tilde}),
\begin{align} \nonumber
&\limsup_n\left|\frac{1}{n^2}\frac{\sqrt{\pmb\tau}^T}{\sqrt{\tau_{t}}}{\bf Z}_N^T{\bf Z}_N\frac{\sqrt{\pmb\tau}}{\sqrt{\tau_{t}}}\right| \\ \nonumber
&\leq\limsup_n\left\|\frac{1}{n}{\bf Z}_N^T{\bf Z}_N\right\|\frac{1}{\tau_t}\frac{1}{n}\sum_{i=1}^n\tau_i<\infty.
\end{align}
Thus, $\frac{1}{N}\frac{1}{n^2}\frac{\sqrt{\pmb\tau}^T}{\sqrt{\tau_{t}}}{\bf Z}_N^T{\bf Z}_N\frac{\sqrt{\pmb\tau}}{\sqrt{\tau_{t}}}\stackrel{\rm a.s.}\longrightarrow0$.

Since the last three terms on the RHS of (\ref{eq:z_tilde}) vanish with large $n$, we obtain $\frac{1}{N}\left|\left\|\tilde{\bf z}_t^2\right\|-\left\|{\bf z}_t^2\right\|\right|\stackrel{\rm a.s.}\longrightarrow0.$
Therefore, as $\frac{1}{N}\|{\bf z}_{t}\|^2\stackrel{\rm a.s.}\longrightarrow\kappa$, we obtain $\frac{1}{N}\|\tilde{\bf z}_{t}\|^2\stackrel{\rm a.s.}\longrightarrow\kappa$ and the convergence (\ref{eq:estgamma}) unfolds.

\section{Proof of Theorem \ref{thrm:estimate}} \label{appx_prfth2}
According to Lemma \ref{lemma:FR2} and Lemma \ref{lemma:Fran}, in which we set ${\bf A}_N={\bf 0}$, $w=\rho$ and ${\bf a}_N=\frac{1}{\sqrt{N}}{\bf 1}_N$, the convergence (\ref{eq:asympestNMR}) at the top of the next page holds.
\begin{figure*}[!t]
\small
\begin{align}\label{eq:asympestNMR}
\sup_{\rho\in\mathcal{R}_\varepsilon}\left|\frac{1}{N}{\bf 1}_N^T\hat{\bf C}_{\rm ST}^{-1}\left(\hat{\bf C}_{\rm ST}-\rho{\bf I}_N\right)\hat{\bf C}_{\rm ST}^{-1}{\bf 1}_N-\frac{k\gamma}{(\gamma+e_N(\rho) k)^2}\frac{\frac{1}{N}{\bf 1}_N^T{\bf C}_N\left(\frac{k}{(\gamma+e_N(\rho)k)}{\bf C}_N+\rho{\bf I}_N\right)^{-2}{\bf 1}_N}{1-\frac{k^2}{(\gamma+e_N(\rho)k)^2}\dfrac{1}{n}{\rm tr}\left[{\bf C}_N^2\left(\frac{k}{(\gamma+e_N(\rho)k)}{\bf C}_N+\rho{\bf I}_N\right)^{-2}\right]}\right|\stackrel{\rm a.s.}\longrightarrow0.
\end{align}
\hrule
\normalsize
\end{figure*}

As $\left|e_N(\rho)-c\gamma\right|\stackrel{\rm a.s.}\longrightarrow0$ when $N, n\rightarrow\infty$, we substitute $c\gamma$ for $e_N(\rho)$ in (\ref{eq:asympestNMR}), giving
\begin{align} \nonumber
&\sup_{\rho\in\mathcal{R}_\varepsilon}\left|\frac{1}{N}{\bf 1}_N^T\hat{\bf C}_{\rm ST}^{-1}\left(\hat{\bf C}_{\rm ST}\!-\!\rho{\bf I}_N\right)\hat{\bf C}_{\rm ST}^{-1}{\bf 1}_N-\frac{(1\!-\!\rho)\!-\!(1\!-\!\rho)^2c}{\gamma}\right. \\ \nonumber
&\left.\times\frac{\gamma^2}{\gamma^2-\beta(1-\rho)^2}\dfrac{1}{N}{\bf 1}_N^T{\bf C}_N\!\left(\!\frac{1-\rho}{\gamma}{\bf C}_N+\rho{\bf I}_N\!\right)^{-2}\!\!\!\!{\bf 1}_N\right|\stackrel{\rm a.s.}\longrightarrow0.
\end{align}

With respect to the asymptotic equivalence in (\ref{eq:asympNMR}) and upon substituting $\hat{\gamma}_{\rm sc}$ for $\gamma/\kappa$, we obtain
\begin{align}\nonumber
&\sup_{\rho\in\mathcal{R}_\varepsilon}\left|\frac{1}{\kappa N}{\bf 1}_N^T\hat{\bf C}_{\rm ST}^{-1}{\bf C}_N\hat{\bf C}_{\rm ST}^{-1}{\bf 1}_N-\frac{\hat{\gamma}_{\rm sc}}{(1-\rho)-(1-\rho)^2c_N}\right. \\ \nonumber
&\left.\times\frac{1}{N}{\bf 1}_N^T\hat{\bf C}_{\rm ST}^{-1}\left(\hat{\bf C}_{\rm ST}-\rho{\bf I}_N\right)\hat{\bf C}_{\rm ST}^{-1}{\bf 1}_N\right|\stackrel{\rm a.s.}\longrightarrow0.
\end{align}
Thus we obtain the consistent estimator of $\frac{1}{\kappa}\sigma^2(\hat{\bf h}_{\rm ST}(\rho))$ in Theorem \ref{thrm:estimate}.

\section{Proof of Corollary \ref{cor:minrho}} \label{appx_prfcor1}
According to Theorem \ref{thrm:estimate}, we have
\begin{align} \nonumber
\sup_{\rho\in{{\mathcal{R}}_{\varepsilon}}}\left|\hat{\sigma}_{\rm sc}^2(\rho)-\frac{1}{\kappa}\sigma^2(\hat{\bf h}_{\rm ST}(\rho))\right|\stackrel{\rm a.s.}\longrightarrow0.
\end{align}
Then, the following holds true
\begin{align} \nonumber
\hat{\sigma}_{\rm sc}^2(\rho^o)&\leq\hat{\sigma}_{\rm sc}^2(\hat{\bf h}_{\rm ST}(\rho^*)) \\ \nonumber
\frac{1}{\kappa}\sigma^2(\hat{\bf h}_{\rm ST}(\rho^*))&\leq\frac{1}{\kappa}\sigma^2(\hat{\bf h}_{\rm ST}(\rho^o)) \\
\nonumber
\hat{\sigma}_{\rm sc}^2(\rho^o)-\frac{1}{\kappa}\sigma^2(\hat{\bf h}_{\rm ST}(\rho^o))&\leq\sup_{\rho\in{{\mathcal{R}}_{\varepsilon}}}\left|\hat{\sigma}_{\rm sc}^2(\rho)-\frac{1}{\kappa}\sigma^2(\hat{\bf h}_{\rm ST}(\rho))\right| \\ \nonumber
&\stackrel{\rm a.s.}\longrightarrow0 \\ \nonumber
\hat{\sigma}_{\rm sc}^2(\rho^*)-\frac{1}{\kappa}\sigma^2(\hat{\bf h}_{\rm ST}(\rho^*))
&\leq\sup_{\rho\in{{\mathcal{R}}_{\varepsilon}}}\left|\hat{\sigma}_{\rm sc}^2(\rho)-\frac{1}{\kappa}\sigma^2(\hat{\bf h}_{\rm ST}(\rho))\right| \\ \nonumber
&\stackrel{\rm a.s.}\longrightarrow0.
\end{align}
These four relations together ensure that
\begin{align} \nonumber
|\sigma^2(\hat{\bf h}_{\rm ST}({\rho}^o))-\sigma^2(\hat{\bf h}_{\rm ST}(\rho^*))|\stackrel{\rm a.s.}\longrightarrow0.
\end{align}

\section{Proof of Proposition \ref{prop: 1}} \label{prof: prop1}
Denote
\begin{align}\nonumber
\left|\widetilde{m}_{N,j}-m_{N,j}\right|=|-A-B+C-D|,
\end{align}
where $1\leq j\leq n$ and
\small
\begin{align}\nonumber
A&\triangleq\frac{1}{N}\frac{1}{n}\frac{\sqrt{\pmb\tau}^T}{\sqrt{\tau_j}}{\bf Z}_N^T{\bf M}_{N,j}{\bf z}_j \\ \nonumber
B&\triangleq\frac{1}{N}{\bf z}_j^T{\bf M}_{N,j}\frac{1}{n}{\bf Z}_N\frac{\sqrt{\pmb\tau}}{\sqrt{\tau_j}} \\ \nonumber
C&\triangleq\frac{1}{N}\frac{1}{n}\frac{\sqrt{\pmb\tau}^T}{\sqrt{\tau_j}}{\bf Z}_N^T{\bf M}_{N,j}\frac{1}{n}{\bf Z}_N\frac{\sqrt{\pmb\tau}}{\sqrt{\tau_j}} \\ \nonumber
D&\triangleq\frac{1}{N}\tilde{\bf z}_j^T\widetilde{\bf M}_{N,j}\frac{1-\rho_n}{1-(1-\rho_n)c_N}\frac{1}{n}\left(\frac{1}{n^2}\sum_{t\neq j}{\bf Z}_N\frac{\sqrt{\pmb\tau}}{\sqrt{\tau_t}}\frac{\sqrt{\pmb\tau}}{\sqrt{\tau_t}}^T{\bf Z}_N^T\right. \\ \nonumber
&~~\left.-\frac{1}{n}\sum_{t\neq j}{\bf z}_t\frac{\sqrt{\pmb\tau}}{\sqrt{\tau_t}}^T{\bf Z}_N^T-\frac{1}{n}\sum_{t\neq j}{\bf Z}_N\frac{\sqrt{\pmb\tau}}{\sqrt{\tau_t}}{\
\bf z}_t^T\right)
{\bf M}_{N,j}\tilde{\bf z}_j.
\end{align}
\normalsize
We wish to prove that
\begin{align} \label{eq:proveprop1}
E[\left|\widetilde{m}_{N,j}-m_{N,j}\right|^p]\leq\frac{K_p}{N^p}
\end{align}
for some constants $p\geq1$, where $K_p$ depends on $p$ but not on $N$. Then, taking $p\geq2$, along with the union bound, the Markov inequality, and the Borel-Cantelli lemma, completes the proof of Proposition \ref{prop: 1}.

Using the Minkowski inequality, we have
\begin{align} \nonumber
&E\left[\left|\widetilde{m}_{N,j}-m_{N,j}\right|^p\right]\leq \\ \nonumber
&(E^{1/p}[|A|^p]+E^{1/p}[|B|^p]+E^{1/p}[|C|^p]+E^{1/p}[|D|^p])^p.
\end{align}
Thus, to prove (\ref{eq:proveprop1}), it is enough to show that $E[|A|^p]\leq\frac{K_{pA}}{N^p}$, $E[|B|^p]\leq\frac{K_{pB}}{N^p}$, $E[|C|^p]\leq\frac{K_{pC}}{N^p}$ and $E[|D|^p]\leq\frac{K_{pD}}{N^p}$.

\subsection{Moments of $|A|$, $|B|$ and $|C|$} \label{sec:A}
Start by noting that
\begin{align} \nonumber
&E[|A|^p]\!=\!\frac{1}{N^p}\frac{1}{n^p}E\!\left[\left|{\bf z}_j^T{\bf M}_{N,j}{\bf Z}_N\frac{\sqrt{\pmb\tau}}{\sqrt{\tau_j}}\frac{\sqrt{\pmb\tau}^T}{\sqrt{\tau_j}}{\bf Z}_N^T{\bf M}_{N,j}{\bf z}_j\right|^{p/2}\right] \\ \nonumber
&=\frac{1}{N^p}\frac{1}{n^p}E\left[\left|{\bf z}_j^T{\bf M}_{N,j}\left({\bf z}_j+{\bf Z}_N^{(j)}\frac{\sqrt{{\pmb\tau}^{(j)}}}{\sqrt{\tau_j}}\right) \right.\right.\\ \nonumber
&~~~~~~~~~~~~~~~~~~~~~~~~\left.\left.\times\left({\bf z}_j+{\bf Z}_N^{(j)}\frac{\sqrt{{\pmb\tau}^{(j)}}}{\sqrt{\tau_j}}\right)^T{\bf M}_{N,j}{\bf z}_j \right|^{p/2}\right] \\ \nonumber
&=\frac{1}{N^p}\frac{1}{n^p}E\left[\left|{\bf z}_j^T{\bf M}_{N,j}\left({\bf z}_j{\bf z}_j^T+{\bf z}_j\frac{\sqrt{{\pmb\tau}^{(j)}}^T}{\sqrt{\tau_j}}{\bf Z}_N^{(j)T}\right.\right.\right. \\ \nonumber
&\left.\left.\left.+{\bf Z}_N^{(j)}\frac{\sqrt{{\pmb\tau}^{(j)}}}{\sqrt{\tau_j}}{\bf z}_j^T+{\bf Z}_N^{(j)}\frac{\sqrt{{\pmb\tau}^{(j)}}}{\sqrt{\tau_j}}\frac{\sqrt{{\pmb\tau}^{(j)}}^T}{\sqrt{\tau_j}}{\bf Z}_N^{(j)T}\right){\bf M}_{N,j}{\bf z}_j\right|^{p/2}\right] \\ \nonumber
&\stackrel{(a)}\leq A_1+A_2+A_3+A_4,
\end{align}
where $(a)$ follows from Jensen's inequality and
\begin{align} \nonumber
A_1&=\frac{4^{p/2-1}}{N^pn^p}E\left[\left|{\bf z}_j^T{\bf M}_{N,j}{\bf z}_j{\bf z}_j^T{\bf M}_{N,j}{\bf z}_j\right|^{p/2}\right] \\ \nonumber
A_2&=\frac{4^{p/2-1}}{N^pn^p}E\left[\left|{\bf z}_j^T{\bf M}_{N,j}{\bf Z}_N^{(j)}\frac{\sqrt{{\pmb\tau}^{(j)}}}{\sqrt{\tau_j}}\right.\right. \\ \nonumber
&\left.\left.~~~~~~~~~~~~~~~~~~\times\frac{\sqrt{{\pmb\tau}^{(j)}}^T}{\sqrt{\tau_j}}{\bf Z}_N^{(j)T}{\bf M}_{N,j}{\bf z}_j\right|^{p/2}\right] \\ \nonumber
A_3&=\frac{4^{p/2-1}}{N^pn^p}E\left[\left|{\bf z}_j^T{\bf M}_{N,j}{\bf Z}_N^{(j)}\frac{\sqrt{{\pmb\tau}^{(j)}}}{\sqrt{\tau_j}}{\bf z}_j^T{\bf M}_{N,j}{\bf z}_j\right|^{p/2}\right] \\ \nonumber
A_4&=\frac{4^{p/2-1}}{N^pn^p}E\left[\left|{\bf z}_j^T{\bf M}_{N,j}{\bf z}_j\frac{\sqrt{{\pmb\tau}^{(j)}}^T}{\sqrt{\tau_j}}{\bf Z}_N^{(j)T}{\bf M}_{N,j}{\bf z}_j\right|^{p/2}\right].
\end{align}

For term $A_1$,
\begin{align} \nonumber
{A}_1&=\frac{1}{N^p}\frac{1}{n^p}\cdot4^{p/2-1}E\left[\left|{\bf z}_j^T{\bf M}_{N,j}{\bf z}_j\right|^p\right] \\ \nonumber
&\leq\frac{1}{N^p}\frac{1}{n^p}\cdot4^{p/2-1}E\left[\|{\bf y}_j\|^{2p}\|{\bf M}_{N,j}\|^p\|\|{\bf C}_N\|^p\right].
\end{align}
Note also that
\begin{align} \nonumber
\|{\bf M}_{N,j}\|^p\leq\frac{1}{({\gamma}(\rho_n)+\ell)^p\rho_n^p},
\end{align}
and by Minkowsky's inequality,
\begin{align}\nonumber
E[\|{\bf y}_j\|^{2p}]=E(\sum_{i=1}^N y_{i,j}^2)^p
\leq N^{p}E|y_{1,j}|^{2p}\leq K_pN^p.
\end{align}
Thus
\begin{align} \nonumber
A_1\leq\frac{1}{N^p}\frac{K_p{\|{\bf C}_N\|}^p4^{p/2-1}c_N^p}{({\gamma}(\rho_n)+\ell)^p\rho_n^p}\leq\frac{K_{pA_1}}{N^p}.
\end{align}

Now consider $A_2$:
\small
\begin{align}\nonumber
A_2&\stackrel{(a)}\leq\frac{2^{3p/2-3}}{N^pn^p}\left(E\left[\left|{\bf z}_j^T{\bf Q}_N{\bf z}_j-{\rm tr}\left({\bf Q}_N\right)\right|^{p/2}\right]+E\left[\left|{\rm tr}\left({\bf Q}_N\right)\right|^{p/2}\right]\right) \\ \nonumber
&\stackrel{(b)}\leq\frac{K_p}{N^pn^p}E\left[\left(E^{p/4}|z_{1,j}|^4{\rm tr}[{\bf Q}_N{\bf Q}_N^T]\right)^{p/4} \right.\\ \nonumber
&\left.~~~~~~~~~~~~~~~~~+E|z_{1,j}|^p{\rm tr}\left[({\bf Q}_N{\bf Q}_N^T)^{p/4}\right]+E\left|{\rm tr}\left({\bf Q}_N\right)\right|^{p/2}\right] \\ \nonumber
&=\frac{K_p}{N^pn^p}\left(E^{p/4}|z_{1,j}|^4+E|z_{1,j}|^p+1\right)E\|{\bf M}_{N,j}{\bf Z}_N^{(j)}\frac{\sqrt{{\pmb\tau}^{(j)}}}{\sqrt{\tau_j}}\|^p \\ \nonumber
&\leq\frac{K_p}{N^pn^p}\left(E^{p/4}|z_{1,j}|^4+E|z_{1,j}|^p+1\right) \\ \nonumber
&~~~~~~~~~~~~~~~~~~~\times E\left(\|{\bf M}_{N,j}\|^p\|{\bf C}_N\|^{p/2}\left\|{\bf Y}_N^{(j)}\frac{\sqrt{{\pmb\tau}^{(j)}}}{\sqrt{\tau_j}}\right\|^p\right)  \\ \nonumber
&\leq\frac{1}{N^p}\frac{K_pK_{{\bf C}_N}^{p/2}\!\left(\!E^{p/4}|z_{1,j}|^4+E|z_{1,j}|^p+1\!\right)\!}{({\gamma}(\rho_n)+\ell)^p\rho_n^p}E\!\left\|\!\frac{1}{n}{\bf Y}_N^{(j)}\frac{\sqrt{{\pmb\tau}^{(j)}}}{\sqrt{\tau_j}}\!\right\|^p \\ \nonumber
&\leq K_{pA_2}/N^p,
\end{align}
\normalsize
where ${\bf Q}_N={\bf M}_{N,j}{\bf Z}_N^{(j)}\frac{\sqrt{{\pmb\tau}^{(j)}}}{\sqrt{\tau_j}}\frac{\sqrt{{\pmb\tau}^{(j)}}^T}{\sqrt{\tau_j}}{\bf Z}_N^{(j)T}{\bf M}_{N,j}$, $(a)$ follows from Jensen's inequality and $(b)$ follows from the trace lemma \cite[Lemma B.26]{bai2009spectral}.

For $A_3$,
\small
\begin{align} \nonumber
A_3\!\leq\!\frac{4^{p/2-1}}{N^pn^p}E^{1/2}\!\left[\left|{\bf z}_j^T{\bf M}_{N,j}{\bf Z}_N^{(j)}\frac{\sqrt{{\pmb\tau}^{(j)}}}{\sqrt{\tau_j}}\right|^p\right]\!E^{1/2}\left[\left|{\bf z}_j^T{\bf M}_{N,j}{\bf z}_j\right|^{p}\right]\!.
\end{align}
\normalsize

As we have $A_1\leq K_{pA_1}/N^p$ and $A_2\leq{K}_{pA_2}/N^p$, we obtain $A_3\leq K_{pA_3}/N^p$. Following the same reasoning as for $A_3$, we also get $A_4\leq K_{pA_4}/N^p$.

Therefore, we obtain
\begin{align} \nonumber
E[|A|^p]\leq A_1+A_2+A_3+A_4\leq K_{pA}/N^p.
\end{align}

The same reasoning holds for $E[|B|^p]$, giving $E[|B|^p]\leq K_{pB}/N^p$.
\begin{figure*}[!t]
\begin{align}\nonumber
&D_1=\frac{1}{N}\left({\bf z}_j-\frac{1}{n}{\bf Z}_N\frac{\sqrt{\pmb\tau}}{\sqrt{\tau_j}}\right)^T\widetilde{\bf M}_{N,j}\frac{1-\rho_n}{1-(1-\rho_n)c_N}\frac{1}{n}\left(\frac{1}{n^2}\frac{1}{\sqrt{\pmb\tau}}^T\frac{1}{\sqrt{\pmb\tau}}{\bf Z}_N\sqrt{\pmb\tau}\sqrt{\pmb\tau}^T{\bf Z}_N^T\right){\bf M}_{N,j}\left({\bf z}_j-\frac{1}{n}{\bf Z}_N\frac{\sqrt{\pmb\tau}}{\sqrt{\tau_j}}\right) \\ \nonumber
&D_2=\frac{1}{N}\left({\bf z}_j-\frac{1}{n}{\bf Z}_N\frac{\sqrt{\pmb\tau}}{\sqrt{\tau_j}}\right)^T\widetilde{\bf M}_{N,j}\frac{1-\rho_n}{1-(1-\rho_n)c_N}\frac{1}{n}\left(-\frac{1}{n}{\bf Z}_N\frac{1}{\sqrt{\pmb\tau}}\sqrt{\pmb\tau}^T{\bf Z}_N^T\right){\bf M}_{N,j}\left({\bf z}_j-\frac{1}{n}{\bf Z}_N\frac{\sqrt{\pmb\tau}}{\sqrt{\tau_j}}\right) \\ \nonumber
&D_3=\frac{1}{N}\left({\bf z}_j-\frac{1}{n}{\bf Z}_N\frac{\sqrt{\pmb\tau}}{\sqrt{\tau_j}}\right)^T\widetilde{\bf M}_{N,j}\frac{1-\rho_n}{1-(1-\rho_n)c_N}\frac{1}{n}\left(-\frac{1}{n}{\bf Z}_N\sqrt{\pmb\tau}\frac{1}{\sqrt{\pmb\tau}}^T{\bf Z}_N^T\right){\bf M}_{N,j}\left({\bf z}_j-\frac{1}{n}{\bf Z}_N\frac{\sqrt{\pmb\tau}}{\sqrt{\tau_j}}\right).
\end{align}
\hrule
\end{figure*}
\begin{figure*}[b]
\hrule
\begin{align} \nonumber
\!{\bf E}&\!=\!{\bf Z}_N^{(j)}{{\bf Z}_N^{(j)}}^T\!\!\!\!-\!\frac{1}{n}{\bf Z}_N^{(j)}\!\frac{1}{\sqrt{{\pmb\tau}^{(j)}}}\!\left(\!{\bf Z}_N^{(j)}\!\sqrt{{\pmb\tau}^{(j)}}\!\right)^T\!\!\!\!-\!\frac{1}{n}{\bf Z}_N^{(j)}\!\sqrt{{\pmb\tau}^{(j)}}\!\left(\!{\bf Z}_N^{(j)}\!\frac{1}{\sqrt{\pmb\tau}}\!\right)^T+\frac{1}{n^2}\left(\frac{1}{\sqrt{{\pmb\tau}^{(j)}}}\right)^T\frac{1}{\sqrt{{\pmb\tau}^{(j)}}}{\bf Z}_N^{(j)}\sqrt{{\pmb\tau}^{(j)}}\left({\bf Z}_N^{(j)}\sqrt{{\pmb\tau}^{(j)}}
\right)^T \\ \nonumber
{\bf F}&=-\frac{1}{n}{\bf Z}_N^{(j)}\frac{1}{\sqrt{{\pmb\tau}^{(j)}}}{\sqrt{\tau_j}}{\bf z}_j^T-\frac{1}{n}{\sqrt{\tau_j}}{\bf z}_j\left({\bf Z}_N^{(j)}\frac{1}{\sqrt{{\pmb\tau}^{(j)}}}\right)^T+\!\frac{\tau_j}{n^2}\!\left(\!\!\frac{1}{\sqrt{{\pmb\tau}^{(j)}}}\!\right)^T\!\!\!\!\!\frac{1}{\sqrt{{\pmb\tau}^{(j)}}}{\bf z}_j{\bf z}_j^T\!+\!\frac{1}{n^2}\!\left(\!\!\frac{1}{\sqrt{{\pmb\tau}^{(j)}}}\!\right)^T\!\!\!\!\!\frac{1}{\sqrt{{\pmb\tau}^{(j)}}}{\bf Z}_N^{(j)}\sqrt{{\pmb\tau}^{(j)}}\sqrt{\tau_j}{\bf z}_j^T  \\ \nonumber
&~~~+\frac{1}{n^2}\left(\frac{1}{\sqrt{{\pmb\tau}^{(j)}}}\right)^T\!\!\!\!\frac{1}{\sqrt{{\pmb\tau}^{(j)}}}\!\sqrt{\tau_j}{\bf z}_j\left({\bf Z}_N^{(j)}\sqrt{{\pmb\tau}^{(j)}}\right)^T.
\end{align}
\end{figure*}
As for the moments of $|C|$, it is similar to how we dealt with $A_1$:
\begin{align} \nonumber
E[|C|^p]&\leq\frac{1}{N^p}E\left[\left\|\frac{1}{n}{\bf Y}_N\frac{\sqrt{\pmb\tau}}{\sqrt{\tau_j}}\right\|^{2p}\|{\bf C}_N\|^p\|{\bf M}_{N,j}^p\|\right] \\ \nonumber
&\leq\frac{\|{\bf C}_N\|^p}{N^p({\gamma}(\rho_n)+\ell)^p\rho_n^p}E\left[\left\|\frac{1}{n}{\bf Y}_N\frac{\sqrt{\pmb\tau}}{\sqrt{\tau_j}}\right\|^{2p}\right] \\ \nonumber
&\leq \frac{K_{pC}}{N^p}.
\end{align}

\subsection{Moments of $|D|$}
We denote $\frac{1}{\sqrt{\pmb\tau}}=(\frac{1}{\sqrt{\tau_1}},...,\frac{1}{\sqrt{\tau_n}})$ and rewrite $D$ as
\begin{align}\nonumber
D=D_1+D_2+D_3
\end{align}
with $D_1$, $D_2$ and $D_3$ at the top of the page.
We aim to prove $E[|D|^p]\leq K_{pD}/N^p$, which is achieved by proving $E[|D_1|^p]\leq K_{pD_1}/N^p$, $E[|D_2|^p]\leq K_{pD_2}/N^p$ and $E[|D_3|^p]\leq K_{pD_3}/N^p$.

Let's first analyze $D_1$ with the analysis of $D_2$ and $D_3$ following similarly. We obtain
\begin{align}\nonumber
&E[|D_1|^p] \\ \nonumber
&\stackrel{(a)}\leq\frac{1}{N^p}\left(\frac{1-\rho_n}{1-(1-\rho_n)c_N}\right)^p
E^{1/2}[|D_{1a}|^{2p}]E^{1/2}[|D_{1b}|^{2p}] \\ \nonumber
&\stackrel{(b)}\leq\frac{1}{N^p}\left(\frac{1-\rho_n}{1-(1-\rho_n)c_N}\right)^p \\  \label{eq:D_1} &\times (E^{1/{2p}}[|D_{1c}|^{2p}]+E^{1/{2p}}[|D_{1d}|^{2p}])^pE^{1/2}[|D_{1b}|^{2p}]
\end{align}
where $(a)$ follows from the Cauchy-Schwarz inequality, $(b)$ follows from Minkowsky's inequality, and
\begin{align} \nonumber
D_{1a}&=\left({\bf z}_j-\frac{1}{n}{\bf Z}_N\frac{\sqrt{\pmb\tau}}{\sqrt{\tau_j}}\right)^T\widetilde{\bf M}_{N,j}\frac{1}{n^2}\frac{1}{\sqrt{\pmb\tau}}^T\frac{1}{\sqrt{\pmb\tau}}{\bf Z}_N\sqrt{\pmb\tau} \\ \nonumber
D_{1b}&=\frac{1}{n}\sqrt{\pmb\tau}^T{\bf Z}_N^T{\bf M}_{N,j}\left({\bf z}_j-\frac{1}{n}{\bf Z}_N\frac{\sqrt{\pmb\tau}}{\sqrt{\tau_j}}\right) \\ \nonumber
D_{1c}&={\bf z}_j^T\widetilde{\bf M}_{N,j}\frac{1}{n^2}\frac{1}{\sqrt{\pmb\tau}}^T\frac{1}{\sqrt{\pmb\tau}}{\bf Z}_N\sqrt{\pmb\tau} \\ \nonumber
D_{1d}&=\left(\frac{1}{n}{\bf Z}_N\frac{\sqrt{\pmb\tau}}{\sqrt{\tau_j}}\right)^T\widetilde{\bf M}_{N,j}
\frac{1}{n^2}\frac{1}{\sqrt{\pmb\tau}}^T\frac{1}{\sqrt{\pmb\tau}}{\bf Z}_N\sqrt{\pmb\tau}.
\end{align}
\normalsize
Our aim is to prove that $E[|D_{1b}|^{2p}]\leq K_{pb}$, $E[|D_{1c}|^{2p}]\leq K_{pc}$ and $E[|D_{1d}|^{2p}]\leq K_{pd}$.
Following the analysis of $E[|A|^p]$ and $E[|C|^p]$, we obtain $E[|D_{1b}|^{2p}]\leq K_{pb}$.
For $E[|D_{1d}|^{2p}]$, we have
\begin{align}\nonumber
&E[|D_{1d}|^{2p}] \\ \nonumber
&\leq\frac{1}{n^{2p}\tau_j^p}\left\|\frac{1}{\sqrt{\pmb\tau}}\right\|^{4p}E\left[\left\|\widetilde{\bf M}_{N,j}\right\|^{2p}\|{\bf C}_N\|^{2p}\left\|\frac{1}{n}{\bf Y}_N\sqrt{\pmb\tau}\right\|^{4p}\right] \\ \nonumber
&\leq\frac{\|{\bf C}_N\|^{2p}}{n^{2p}\tau_j^p({\gamma}(\rho_n)+\ell)^{2p}\rho_n^{2p}}\left\|\frac{1}{\sqrt{\pmb\tau}}\right\|^{4p}E\left\|\frac{1}{n}{\bf Y}_N\sqrt{\pmb\tau}\right\|^{4p} \\ \nonumber
&\leq K_{pd}.
\end{align}

Let us now establish the inequality for $D_{1c}$. We can see that ${\bf z}_j$ is not independent of $\widetilde{\bf M}_{N,j}$, thus we cannot follow the same procedure as for our analysis of $A$ to determine the order of $E[|D_{1c}|^{2p}]$. Instead, we divide $\widetilde{\bf M}_{N,j}$ into two parts, one that is independent of ${\bf z}_j$ and the other the remainder.

 We first write $\sum_{t\neq j}\tilde{\bf z}_{t}\tilde{\bf z}_{t}^T={\bf E}+{\bf F}$,
where ${\bf E}$ and ${\bf F}$ are defined at the bottom of the page.
Note that ${\bf E}$ is independent of ${\bf z}_j$ and ${\bf F}$ is not. Then $D_{1c}$ can be rewritten as (\ref{eq:D_1c_1}) at the top of the next page.
\begin{figure*}[!t]
\small
\begin{align} \nonumber
D_{1c}=&
{\bf z}_j\left(\frac{1-\rho_n}{1-(1-\rho_n)c_N}\frac{1}{n}{\bf E}+({\gamma}(\rho_n)+\ell)\rho_n{\bf I}_N\right)^{-1}\frac{1}{n^2}\frac{1}{\sqrt{\pmb\tau}}^T\frac{1}{\sqrt{\pmb\tau}}{\bf Z}_N\sqrt{\pmb\tau}+{\bf z}_j\widetilde{\bf M}_{N,j}\left(-\frac{1-\rho_n}{1-(1-\rho_n)c_N}\frac{1}{n}{\bf F}\right)\\ \label{eq:D_1c_1}
&\times\left(\frac{1-\rho_n}{1\!-\!(1-\rho_n)c_N}\frac{1}{n}{\bf E}+({\gamma}(\rho_n)\!+\!\ell)\rho_n{\bf I}_N\right)^{-1}\frac{1}{n^2}\frac{1}{\sqrt{\pmb\tau}}^T\frac{1}{\sqrt{\pmb\tau}}{\bf Z}_N\sqrt{\pmb\tau}.
\end{align}
\normalsize
\hrule
\end{figure*}
Using Jensen's inequality,
\begin{align} \nonumber
E[|D_{1c}|^{2p}]\leq2^{2p-1}\left(E[|G|^{2p}]+E[|H|^{2p}]\right),
\end{align}
where $G$ and $H$ are the two terms on the RHS of (\ref{eq:D_1c_1}).
Next we can use the same technique as used in Appendix \ref{sec:A} to prove that $E[|G|^{2p}]\leq K_{pG}$ and $E[|H|^{2p}]\leq K_{pH}$.
Therefore, we obtain $E[|D_{1c}|^{2p}]\leq K_{pc}$.

Thus far, we have proven that $E[|D_{1b}|^{2p}]\leq K_{pb}$, $E[|D_{1c}|^{2p}]\leq K_{pc}$, and $E[|D_{1d}|^{2p}]\leq K_{pd}$. Coming back to (\ref{eq:D_1}), we obtain $E[|D_1|^p]\leq K_{pD_1}/N^p$.

Following similar arguments to our analysis of $E[|D_1|^p]$, we can also obtain $E[|D_2|^p]\leq K_{pD_2}/N^p$ and $E[|D_3|^p]\leq K_{pD_3}/N^p$. As $D=D_1+D_2+D_3$, by Minkowsky's inequality,
we  obtain $E[|D|^p]\leq K_{pD}/N^p$.
\end{appendices}
\bibliographystyle{IEEEtran}
\bibliography{Cited_2}

\end{document}